\documentclass[jounral]{IEEEtran}
\ifCLASSINFOpdf
\else
\fi

\usepackage{xcolor}
\usepackage{todonotes}
\usepackage{amsmath}
\usepackage{amssymb}
\usepackage{url}
\usepackage{multirow}
\usepackage{pifont}
\graphicspath{{figures/}}
\newcommand{\xmark}{\ding{55}} 

\begin{document}
%
\title{Feature-Oriented IoT Malware Analysis: Extraction, Classification, and Future Directions}

\author{
\IEEEauthorblockN{Zhuoyun Qian}
\and
\IEEEauthorblockN{Hongyi Miao}
\and
\IEEEauthorblockN{Cheng Zhang}
\and
\IEEEauthorblockN{Qin Hu}
\and
\IEEEauthorblockN{Yili Jiang}
\and
\IEEEauthorblockN{Jiaqi Huang}
\and
\IEEEauthorblockN{Fangtian Zhong}
}

		\author{Zhuoyun Qian,
			 Hongyi Miao,
        Cheng Zhang,
			Qin Hu,~\IEEEmembership{Member,~IEEE,}
           Yili Jiang, ~\IEEEmembership{Member,~IEEE,}
           Jiaqi Huang, ~\IEEEmembership{Member,~IEEE,}
  and~Fangtian Zhong,~\IEEEmembership{Member,~IEEE}
			\thanks{Zhuoyun Qian, Hongyi Miao, and Fangtian~Zhong (Corresponding author) are with the Gianforte School of Computing,
				 Montana State University, Bozeman,
				MT 59717, USA. E-mail: \{zhuoyunqian19, hongyimiao10\}@gmail.com, {fangtian.zhong@montana.edu}}
			\thanks{Cheng Zhang is with Department of Computer Information and Decision Management, West Texas A\&M University, Canyon, TX 79016, E-mail: czhang@wtamu.edu}
            \thanks{Qin~Hu and Yili~Jiang are with the Department of Computer Science, Georgia State University, Atlanta, GA 30303, USA. E-mail: \{qhu, yjiang27\}@gsu.edu}
                \thanks{Jiaqi~Huang is with the Department of Computer Science and Cybersecurity, University of Central Missouri, Warrensburg, MO 64093, USA. E-mail: jhuang@ucmo.edu}
        }


%


\maketitle

\begin{abstract}

As Internet of Things (IoT) devices continue to proliferate, their reliability is increasingly undermined by security concerns. Many attackers exploit malware in IoT devices to launch attacks, leading to large-scale data breaches or device damage. Consequently, malware analysis has become particularly important, which can not only help us study existing malware statically, classify them, and devise unified defense strategies, but also enable the detection of malicious program behaviors and network flows, thereby preventing potential attacks. However, malware analysis has never been a straightforward task. Attackers are growing more sophisticated, disguising their malware with varied techniques as new strains emerge rapidly. Although traditional approaches extract useful features from IoT programs for analysis, they fall short in keeping pace with the ever-evolving landscape of malware.
To address these challenges, this survey systematically reviews and analyzes the key research problems and emerging directions in malware analysis. Specifically, we categorize existing techniques into four groups based on their analytical approaches, discuss the types of features  and the processing methods, and examine the challenges each category faces along with potential solutions. Finally, we compare the strengths and limitations of current techniques, highlight open challenges, and outline promising avenues for future research. This survey provides newcomers with a comprehensive entry point into the field and offers security experts a broader perspective to inspire further exploration in IoT malware detection.

\end{abstract}

\IEEEpeerreviewmaketitle

\section{Introduction}
\label{sec:introduction}

Over the past two decades, Internet of Things (IoT) technologies have been increasingly applied across diverse domains, evolving from smart homes to smart cities, industrial IoT, 5G-enabled edge computing, and more recently, healthcare systems and wearable devices. Today, IoT devices are deeply integrated into both industrial production and personal life, driving rapid market growth. As of August 2025, the global IoT market size has reached USD 53.13 billion and is projected to grow to USD 114.54 billion by 2033. This rapid expansion highlights that IoT technologies have become embedded in critical infrastructures, making their security issues increasingly significant \cite{globalgrowth2025iot, cogniteq2024iot}.

However, due to the limited computing resources of IoT devices, the lack of standardization, and the immaturity of security mechanisms, numerous vulnerabilities have emerged. Attackers can exploit these weaknesses to launch botnet, DDoS, and malware attacks against personal and industrial IoT devices, leading to severe consequences such as information leakage and financial losses.

To mitigate these security threats, researchers have proposed a variety of approaches, including static, dynamic, hybrid, and learning-based analyses. These methods typically detect and defend against attacks or malicious programs in IoT environments by analyzing IoT applications, network traffic among IoT devices, and other behavioral patterns. Broadly, such approaches can be divided into two stages: i) feature extraction and processing, and ii) malware or attack analysis. Among them, the efficiency of feature extraction and the ability to capture more representative features of IoT programs and network behaviors remain important research directions.

To address these threats, researchers have explored static, dynamic, hybrid, and learning-based analysis methods. While previous surveys have contributed valuable insights, they often remain limited in scope. For example, \cite{adil2023uav} investigates the security threats in UAV-assisted IoT networks and summarizes the use of machine learning-based methods to address these threats. Similarly, \cite{hussain2020machine, venkatasubramanian2023iot} review the applications of machine learning-based approaches and federated learning in attack detection and malware analysis. In the same vein, \cite{or2019dynamic} focuses primarily on dynamic analysis techniques, systematically discussing both the techniques themselves and their integration with machine learning for malware detection, while highlighting IoT malware analysis as a future research direction. Distinct from these works, \cite{ye2017survey} provides a systematic exploration of malware detection from the perspective of data mining, covering a variety of feature extraction and selection methods as well as different machine learning-based classification and clustering approaches. In contrast, \cite{gaber2024malware} presents a more comprehensive overview of malware detection, encompassing static and dynamic analysis methods, comparisons of static and dynamic features and their selection techniques, and an in-depth discussion of the strengths and weaknesses of using machine learning. Moreover, \cite{raju2021survey} focuses specifically on IoT malware detection, reviewing different analysis approaches and features, and providing a comprehensive survey of cross-architectural IoT malware threat hunting. Importantly, it also proposes a taxonomy that categorizes IoT malware features into four classes: metric-based, graph/tree-based, sequence-based, and interdependence-based, to address the challenges of future cross-architectural IoT malware detection.  

In contrast to existing surveys, our work addresses the lack of a systematic investigation into feature extraction techniques for IoT malware analysis. To the best of our knowledge, this is the first survey that comprehensively examines the features employed across different analysis methods as well as their corresponding extraction techniques. We provide a broad and in-depth discussion of various features, ranging from single-feature approaches to multi-feature combinations, in the context of IoT malware detection and classification. Furthermore, we compare different extraction methods by analyzing their respective strengths and weaknesses, and we investigate why certain techniques and features have been widely adopted, tracing their evolution over time to reveal insights into their development. In particular, we highlight the application of machine learning techniques based on graph features for IoT malware analysis. Ultimately, our study not only summarizes the distinctions between our survey and prior surveys as shown in Table \ref{tab:survey_comparison}, but also identifies unexplored directions and outlines promising avenues for future research in this field.

The key contributions of this survey are summarized as follows:
\begin{enumerate}
    \item We present a comprehensive and modern feature-oriented overview of IoT malware analysis, covering diverse features and feature extraction techniques. Our survey systematically examines static, dynamic, hybrid, and graph learning-based features for malware detection and classification.
    
    \item We investigate various feature extraction, selection, and processing techniques, and compare the use of different feature categories. Special emphasis is placed on their differences, strengths, and limitations in different application scenarios. 
    
    \item We provide an in-depth discussion on both well-established research directions and emerging challenges in IoT malware analysis. The survey concludes by outlining future opportunities for leveraging different features in malware detection and classification.
    
    \item To ensure comprehensiveness, we developed an automated Python script to retrieve relevant articles from ACM, IEEE, Springer, and Google Scholar. For each surveyed paper, we verified dataset and code availability, categorizing them as private, partially public, or public. In our reviewed articles, 47.8\% did not disclose their datasets, 20.3\% partially disclosed them, and only 31.9\% fully provided public access. Regarding dataset sources, the majority of studies relied on a limited set of well-established benchmarks, with VirusShare, Drebin, IoT-23, Bot-IoT, and VirusTotal together accounting for approximately 64.5\% of all dataset usage. All collected data and our scripts are publicly released at \url{https://github.com/QXinHan/IoTMalware.git}.
\end{enumerate}

In the preceding section, we provided a brief introduction to IoT security, summarized and reviewed recent related surveys, highlighted their limitations, and presented our own contributions. The remainder of this survey is organized as follows. Section~\ref{sec:background} outlines common IoT attacks and reviews widely adopted feature extraction techniques in IoT malware analysis, providing essential background knowledge for readers entering this field. Section~\ref{sec:features} then discusses the use of diverse features, extraction techniques, and malware analysis methods for IoT malware analysis from four distinct perspectives. In Section~\ref{sec:opptunities and challenges}, we compare the strengths and weaknesses of different techniques, summarize the lessons learned, and further explore potential future research directions. Finally, Section~\ref{sec:conclusion} concludes the paper.

\begin{figure}[htbp]
    \centering
    \includegraphics[width=\columnwidth]{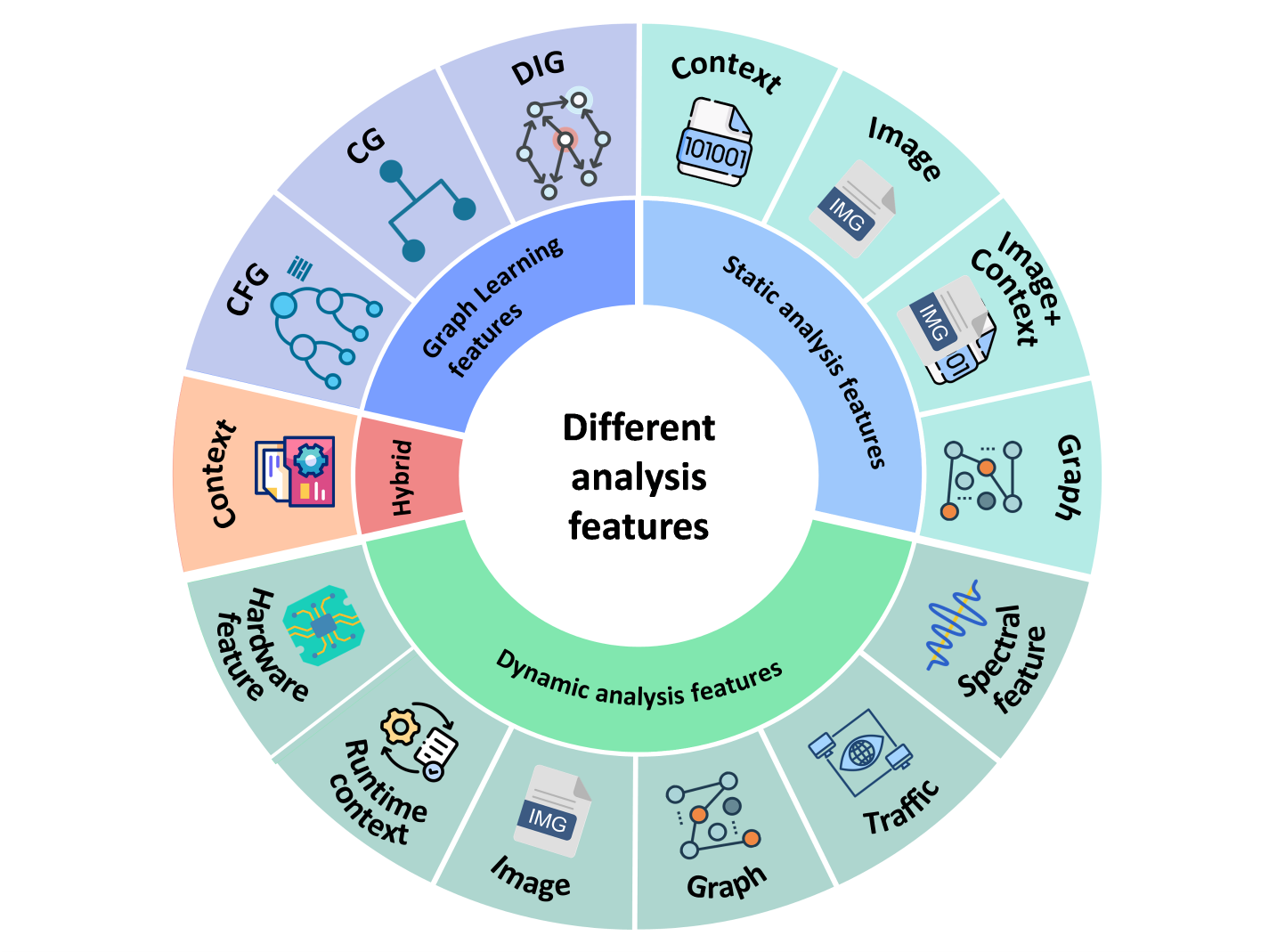}
    \caption{Overview of the survey.}
    \label{fig:Overview_survey}
\end{figure}

\begin{table*}[htbp]
\centering
\caption{Comparison of Existing Related Surveys}
\begin{tabular}{|l|c|c|c|}
\hline
\textbf{Survey} & 
\textbf{Diverse Features} & 
\textbf{Diverse Feature Extraction Techniques} & 
\textbf{IoT Malware Analysis} \\ \hline

\cite{adil2023uav} & 
\xmark & 
\xmark & 
\checkmark \\ \hline

\cite{hussain2020machine} & 
\xmark & 
\xmark & 
\xmark \\ \hline

\cite{venkatasubramanian2023iot} & 
\checkmark & 
\xmark & 
\checkmark \\ \hline

\cite{gaber2024malware} & 
\checkmark & 
\checkmark & 
\xmark \\ \hline

\cite{ye2017survey} & 
\checkmark & 
\checkmark & 
\xmark \\ \hline

\cite{raju2021survey} & 
\checkmark & 
\xmark & 
\checkmark \\ \hline

\cite{or2019dynamic} & 
\xmark & 
\xmark & 
\xmark \\ \hline

Our survey & 
\checkmark &
\checkmark &
\checkmark \\ \hline

\end{tabular}
\label{tab:survey_comparison}
\end{table*}

\section{BACKGROUND}
\label{sec:background}

\subsection{Common IoT Attacks}

Due to the resource-constrained nature of IoT devices, high degree of connectivity, immaturity of corresponding system defenses, architectural diversity, and prevalence of Linux- or Android-based platforms, IoT malware  attacks are mainly concentrated in the following categories:

\begin{enumerate}

\item \textbf{DDoS and Botnet Propagation (e.g., Mirai, Gafgyt, Tsunami):} Attackers compromise large numbers of IoT devices to form botnets, which are then leveraged to launch massive volumes of requests against target servers or networks, causing service outages. The vast number of IoT devices, their frequent exposure to public networks, and weak authentication mechanisms make them easy to compromise and control, enabling attackers to build large-scale botnets and conduct powerful DDoS attacks.

\item \textbf{Cross-architecture ELF Malware (e.g., XorDDOS):} IoT devices operate on heterogeneous hardware architectures (e.g., ARM, MIPS, PPC, x86). Attackers often cross-compile ELF binaries into multiple platform-specific versions, thereby achieving the principle of "write once, run everywhere." Such malware can propagate rapidly across heterogeneous platforms, significantly expanding the attack surface.

\item \textbf{Malicious Android Applications:} Many IoT devices rely on the Android operating system or companion mobile applications for management and interaction, providing attackers with an effective entry point. Malicious applications typically exploit \textit{permission abuse} (e.g., stealing location or SMS data), \textit{sensitive API calls} (e.g., file writing, dynamic loading), or \textit{intent injection} (triggering unauthorized operations) to conduct data theft and device hijacking.

\item \textbf{Counterfeit or Cloned IoT Applications:} The IoT ecosystem includes a large number of third-party applications but lacks unified security standards and development guidelines. This allows attackers to create counterfeit applications that closely mimic legitimate ones. Such applications are nearly indistinguishable from the originals in terms of user interface and functionality, making them difficult for users to detect. Attackers often embed malicious code into these clones, such as adware injection or sensitive data exfiltration.

\item \textbf{Command Injection and Malicious Firmware Updates (e.g., Shell injection, firmware Trojans):} IoT devices typically run Linux-based systems and support remote command execution as well as firmware updates. These features create opportunities for attackers to compromise systems at a fundamental level. For instance, attackers can inject malicious commands via the shell interface or distribute tampered firmware images that implant Trojans or backdoors into devices.

\item \textbf{Data Exfiltration and Side-channel Attacks:} The massive number of IoT devices, coupled with insufficient traffic monitoring and the lack of fine-grained inspection of DNS/HTTP communications, provides cover for data exfiltration. Since devices routinely transmit data to cloud services or vendor servers, attackers can disguise malicious traffic within such legitimate flows, increasing the risk of sensitive data leakage. Moreover, IoT devices are often deployed in physically accessible environments and built with hardware of varying quality, making it easier for attackers to exploit electromagnetic (EM) emissions, power consumption patterns, or implant monitoring devices to perform side-channel attacks.

\end{enumerate}

\subsection{A Summary of Feature Extraction Techniques}

\subsubsection{Static Feature Extraction Techniques}

The method of extracting features directly from binary files without executing the program is referred to as static feature extraction techniques. Commonly used static feature extraction methods can be broadly categorized into four groups. First, disassembly tools (e.g., IDA Pro, Radare2, Object-Dump, Dexdump) or customized parsers are employed to extract features such as opcodes, file attributes, API calls, permissions, and intents from APK, ELF, or PE files. Second, statistical techniques are applied to derive features from opcodes or binary files, including TF–IDF representations, byte histograms, byte entropy histograms, and string statistics. Third, image-based features are constructed by direct byte-to-pixel mapping, Markov transition matrices, or context-aware mappings, which transform executables into grayscale or RGB images. Finally, graph-based features are extracted by constructing control-flow graphs (CFGs) with tools such as Capstone, followed by representation learning methods like Graph2Vec to obtain graph embeddings.

\subsubsection{Dynamic Feature Extraction Techniques}

The approach of obtaining data by executing the program and monitoring the program’s runtime state, the device’s operational state, or the dynamic interactions within the entire IoT network environment, and then extracting features from this data, is referred to as dynamic feature extraction techniques. Common dynamic feature extraction methods can be broadly categorized into several types. First, temporal or causal logic modeling and statistical feature extraction are widely used to derive behavioral patterns of IoT devices from ETB, HPC, EM signals, sensor readings, logs, or communication traces. Second, tools such as CICFlowMeter are employed to parse raw traffic into structured flow-level features. These traffic-derived features are often further refined using statistical techniques, including information gain, correlation analysis, and temporal analysis, to select the most representative attributes of device behavior. Third, signal processing methods such as FFT, STFT, and CWT are applied to electromagnetic or power traces to extract frequency-band, energy, or shape-related characteristics. Fourth, image-based representations are constructed either by directly mapping traffic bytes or by transforming side-channel information into spectrogram-like formats (e.g., CWT, Mel spectrograms, TFR, or HOG), from which semantic features are extracted. Finally, graph-based features are obtained through network-flow or data-flow analysis, capturing the communication patterns among IoT devices.

\subsubsection{Hybrid Feature Extraction Techniques}

The method that combines static and dynamic feature extraction techniques to obtain both static and dynamic features is referred to as hybrid feature extraction techniques. Common hybrid feature extraction approaches combine multiple techniques. First, code analysis is employed to extract association features between events (e.g., trigger–action rules). Static information such as permissions, API calls, and control flow graphs (CFGs) is obtained using reverse-engineering tools like APKTool and Androguard, while additional static features are derived from hash values and signatures. In parallel, dynamic monitoring is used to capture causal and temporal logic features of runtime events. Emulators such as Monkey and UIAutomator are typically adopted to simulate execution, from which runtime logs are collected to extract function call frequencies and behavioral patterns of program logic. By integrating these static and dynamic analysis methods, both static and dynamic features are combined to infer IoT device behaviors.

\subsubsection{Graph Learning Representation Extraction Techniques}

The approach of extracting graph-based features from programs that can be analyzed and inferred by graph learning models is referred to as graph learning representation extraction techniques. Common approaches first construct graph representations such as API graphs, control flow graphs (CFGs), or device-level dynamic interaction graphs using tools including Androguard with Gephi, Radare2, ApkTool, FlowDroid, or AstBuilder. Subsequently, graph-based features are typically extracted through three main categories of techniques:
(i) Graph-theoretic analysis, which computes structural metrics such as degree, betweenness, closeness, eigenvector, and PageRank centralities;
(ii) Sequence-based methods, which transform graphs into serialized features via random walks, $n$-gram extraction, or sensitive API sequence modeling; and
(iii) Subgraph-oriented approaches, which mine frequent subgraph patterns or construct masked subgraphs centered on high-degree nodes, and then encode them into vector representations using methods such as high-dimensional one-hot encoding, graph2vec embeddings, or advanced graph analysis frameworks.

\section{Malware Analysis Across Features}
\label{sec:features}
There are four primary types of features used to detect malware: static analysis, dynamic analysis, hybrid analysis, and graph representation learning. These features collectively enhance the detection of malware. We will present the survey as illustrated in Figure \ref{fig:Overview_survey}.

\subsection{Static Analysis Feature}

\begin{figure*}[htbp]   
    \centering
    \includegraphics[width=0.7\linewidth, height=4cm]{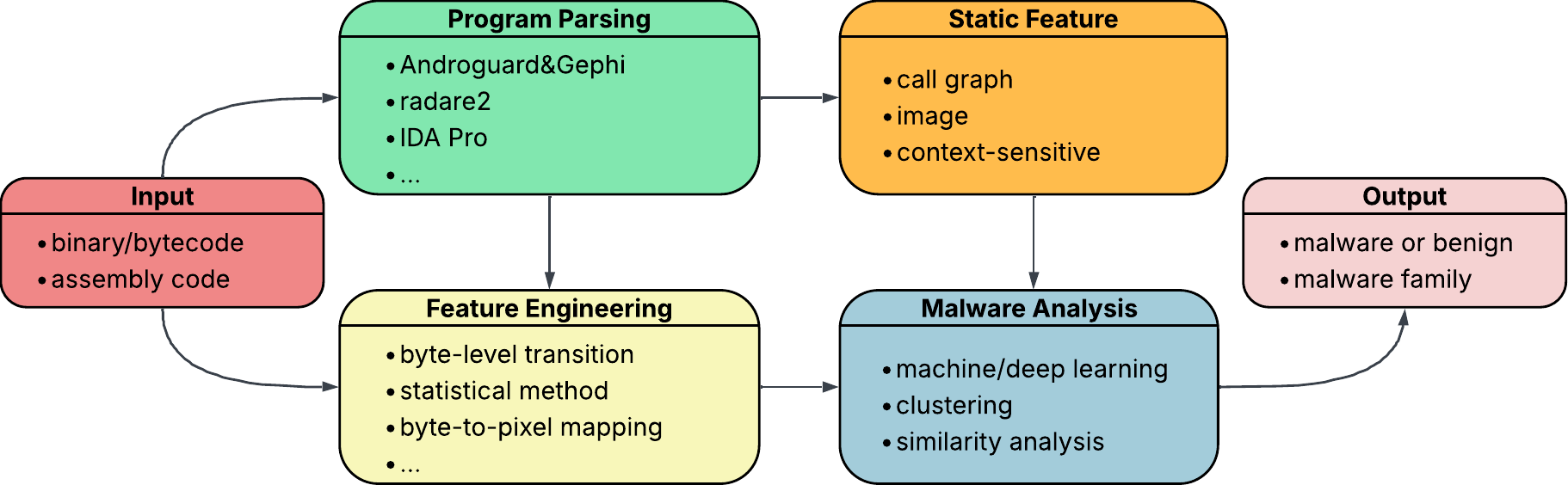}
    \caption{Workflow of static analysis.}
    \label{fig:Static_analysis}
\end{figure*}
The common workflow of static analysis-based malware detection and classification is illustrated in Figure \ref{fig:Static_analysis}. Some methods first process the input binary or bytecode using either customized algorithms or established tools (e.g., \texttt{Radare2}, \texttt{IDA Pro}), which are then subjected to feature engineering to derive three major categories of features: \textit{context-sensitive features}, \textit{images}, and \textit{graph}, or are directly utilized as static features for the next stage. Alternatively, other methods take assembly code (e.g., opcode sequences) as input and extract features during the feature engineering stage through hashing or statistical analysis. Based on these features, malware detection and classification are ultimately performed using four categories of approaches: machine/deep learning-based method, clustering, and similarity analysis.



\subsubsection{Context-Sensitive Feature}

Studies employing this type of feature can be broadly categorized into two groups. The first group are static features extracted from the input executable file using analysis tools, including opcode sequences, temporal–spatial attributes and shell command sequences. 

Several studies follow a multi-stage process, where binaries are first parsed to obtain byte-level or instruction-level sequences, which are then processed through feature engineering to generate feature vectors for malware detection and classification. For instance, SOMRN \cite{xie2024towards} extracts opcode and API call sequences from malware samples and segments them into sliding windows. These windows are subsequently processed using a context-separated bi-directional long short-term memory (CS-BiLSTM) model combined with a Vector Quality Estimator to produce semantic vectors, which are then used for malware classification. Building on a similar idea of leveraging instruction-level semantics, EVOLIoT \cite{dib2022evoliot} disassembles malware binaries with IDA Pro to obtain basic blocks. A customized instruction normalization strategy is then applied to segment these blocks into token sequences, which are fed into a pre-trained (Bidirectional Encoder Representations from Transformers)BERT model under a contrastive learning framework to generate semantically rich representations and enable clustering of unlabeled samples. 

MTHAEL \cite{vasan2020mthael} uses the Object-Dump tool to extract opcode sequences, generates n-gram frequency vectors, and applies two feature selection techniques (Information Gain and an Opcode Dictionary) to obtain more compact representations. These vectors are then fed into an ensemble of CNN and RNN models for detection. In contrast, MFSRC \cite{alaeiyan2020multilabel} leverages IDA Pro to parse opcode sequences, constructs opcode frequency vectors, applies fuzzy transformations to derive low-dimensional relevance vectors, and employs fuzzy clustering with cosine similarity to classify malware families.  The method proposed in \cite{jahromi2020enhanced} follows a simpler pipeline without program parsing because it directly utilizes bytecode or opcode sequences as input. A sliding-window encoding scheme is then applied to generate dense vector embeddings from the sequences, which are subsequently processed by a stacked LSTM with layer-wise pre-training to learn abstract representations. The resulting features are then fed into a classifier to differentiate benign from malicious programs.

The program’s temporal–spatial attributes constitute another important type of feature for representing program behavior. TSDroid \cite{zhang2023tsdroid} first extracts temporal and spatial features from APK files: temporal features are derived from API lifecycle information, while spatial features capture application-level characteristics such as code and package size. These features are integrated into a unified vector representation and clustered to group similar samples. Within each cluster, three pre-trained detection models are used: Drebin (static features + linear SVM), HinDroid (heterogeneous information network + multi-kernel learning), and DroidEvolver (ensemble of online models with weighted voting). The results of these models are fused to perform final malware detection.

ShellCore \cite{alasmary2021shellcore} focuses on extracting shell commands from software samples for malware detection. Given a sample, ShellCore first extracts a shell command sequence and applies two modeling strategies: a term-level model, which transforms the sequence into a bag-of-words and $n$-gram feature space, and a character-level model, where each character is treated as an individual token. Both models convert the shell command sequences into multihot vectors. To reduce dimensionality, ShellCore applies Principal Component Analysis (PCA) to both types of feature vectors. The reduced feature vectors are fed into several malware classifiers, such as Logistic Regression (LR), Random Forest (RF), and Deep Neural Networks (DNN), that can determine whether a program is malicious based on the corresponding feature vector.

The second group adopts a multi-category feature fusion approach, where various types of features are first derived from static analysis outputs using techniques such as N-gram modeling, term frequency–inverse document frequency (TF–IDF), and byte entropy computation. These heterogeneous features are subsequently combined to provide a comprehensive representation of program behavior.

For example, in \cite{li2022imbalanced}, the proposed method takes as input a malware sample consisting of a BYTE file and an ASM file extracts four types of features: (1) byte-level features, including Byte Histogram, Byte Entropy Histogram, and Printable String Statistics; (2) format features, such as field size statistics; (3) TF-IDF vectors derived from strings and assembly sequences using 1- to 3-gram representations; and (4) function-level assembly embeddings generated via Word2Vec. These four types of feature vectors are then processed using three different multimodal fusion strategies to obtain a set of fused representations. The resulting vectors are passed through an Extra Trees Classifier based on Gini importance to select the top-k most informative vectors. Finally, the selected vectors are fed into a pre-trained XGBoost classifier, which outputs the predicted malware type. Cypider \cite{karbab2016cypider} also follows this paradigm: APKs are disassembled with dexdump, and heterogeneous features such as opcode/byte n-grams, API usage, permissions, manifest data, and TF–IDF strings are extracted. These are transformed via feature hashing, and a similarity network is constructed using Locality Sensitive Hashing (LSH) which is finally fed into a One-Class Support Vector Machine (OC-SVM) for malware detection.

Some studies parse executables into higher-level features and feed them directly into classifiers, without additional feature engineering. \cite{tien2020machine} uses Radare2 to disassemble binaries and extracts extracts ELF-level attributes (e.g., instruction set architecture (ISA), file size, and etc.) along with opcode features such as logic operations, memory access and so on. These two feature sets are then independently fed into separate CNN-based models: one trained to distinguish malware from benignware, and the other to classify the detected malware into specific families. Similarly, both Fed-IIoT \cite{taheri2021fediiot} and \cite{deng2023edgebased} extracts three types of static features—API calls, permissions, and intents—from Android IoT applications and embedding them into vector representations for subsequent deep learning-based malware detection. The key difference lies in their feature encoding and utilization strategies. Fed-IIoT encodes these features into a binary vector, where each dimension corresponds to a specific static feature, with a value of 1 indicating the presence of the feature and 0 otherwise. Finally, it employs federated learning and Attention-based Adaptive Adversarial Generative Adversarial Network (A3GAN) with adversarial robustness to detect potential poisoning attacks. In contrast, \cite{deng2023edgebased} directly feeds the three feature types into a TSA-Caps model, which encodes them into a unified multimodal feature vector. This vector is then passed to a pre-trained deep neural network for final classification to determine whether the APK is malicious.

A third group bypasses explicit program parsing and instead directly constructs multimodal statistical representations from input data. Qiao et al. \cite{qiao2021malware} apply CBOW Word2Vec to encode both raw bytes and disassembled instructions into a unified matrix, which is then classified using a Multilayer Perceptron (MLP). REMSF \cite{yu2023remsf} adopts a more complex design by fusing five feature types: byte histograms, byte entropy histograms, string statistics, TF–IDF features, and heterogeneous graphs (HGs). The first four are normalized and concatenated into vectors, while HGs are embedded using the metapath2vec algorithm. These representations are separately processed by four models—Gradient Boosting Decision Trees (GBDT) for byte histograms and string features, a Backpropagation Neural Network (BPNN) for entropy features, a Random Forest for TF–IDF features, and a Support Vector Machine (SVM) with a polynomial kernel for semantic features. Their outputs are then integrated through another GBDT to produce the final detection result. Similarly, Wu et al. \cite{wu2020detection} measure three types of similarity between a given application and verified official ones—TLSH-based whole-app similarity, Minhash-based resource similarity, and motif-based code similarity—and combine them to effectively detect counterfeit/malicious applications.

\subsubsection{Image}

Image-based malware detection methods differ primarily in how the program is transformed into images. Existing approaches can be grouped into four categories. In \cite{zhang2024deep, ullah2023privacy, ravi2023vit4mal}, malware binaries are transformed into grayscale or RGB images through direct byte-to-pixel mapping. Specifically, \cite{zhang2024deep} converts binaries into grayscale images and employs a ResNet50 network to extract feature vectors, which are subsequently converted into binary hash codes; Hamming distances between these codes are then used for clustering malware samples. Similarly, \cite{ravi2023vit4mal} normalizes the RGB images and processes them with a lightweight Vision Transformer (ViT) to obtain deep feature vectors, which are then passed to a multilayer perceptron (MLP) for final classification. Building on this idea, \cite{ullah2023privacy} applies a colormap to further convert grayscale images into color images and classifies them using a federated learning–based 2D CNN framework. 

Instead of raw bytes, disassembled code and metadata are used to construct RGB images. \cite{li2021cnnbased} employs IDA Pro to extract three key types of information, including binary code, assembly instructions, and developer-related strings and then maps them into the R, G, and B channels, respectively to construct an RGB image. The generated RGB image is subsequently processed through a Self-Attention module, which captures global dependencies among pixels, followed by a CNN module that extracts high-dimensional representations. The resulting feature are then passed into a Spatial Pyramid Pooling (SPP) module to unify them into a fixed-size vector representation. Finally, the output vector is fed into a Softmax layer to produce the probability distribution over malware classes.

Some studies combine heterogeneous static features by transforming each into image representations. Mal3S \cite{jeon2024static} constructs five types of images from raw bytes, opcodes, API calls, DLL imports, and visible strings. These are processed by a multi-branch Multi-SPP-net(Multi-Scale Spatial Pyramid Pooling Network), where the byte image is processed independently to generate a dedicated feature vector, while the remaining four images are fused and mapped into another feature vector. Finally, these two feature vectors are passed through a Softmax layer to yield the final malware classification.

Another strategy models byte or opcode transition probabilities and visualizes them as Markov images. Specifically, some approaches treat malware as a byte stream and generate Markov images by modeling the transition patterns of bytes \cite{yuan2022iot,dhanya2023obfuscated}, while others disassemble binaries into opcode sequences and construct Markov images based on opcode transitions \cite{mai2023mobilenet}. More concretely, in \cite{yuan2022iot}, each malware binary is represented as a byte stream, and multiple basic Markov images are constructed by calculating transition frequencies between consecutive byte pairs; these images are then merged into a composite image. In \cite{dhanya2023obfuscated}, the HEXDUMP tool is employed to convert the DEX file extracted from an APK into a hexadecimal sequence, from which state transition probabilities are computed for each byte, aggregated into a $256 \times 256$ Markov matrix, and finally visualized as a grayscale image. In \cite{mai2023mobilenet}, IDA Pro is used to disassemble PE-format binaries into opcode sequences, followed by the n-gram technique to calculate the transition frequencies of consecutive opcodes, producing a 2-gram Markov transition probability matrix that is subsequently converted into a grayscale Markov image. Once the images are constructed, they are fed into a pre-trained Lightweight Convolutional Classification Network (LCCN) or a Convolutional Neural Network (CNN) to extract deep feature vectors and ultimately predict the malware family or type.

\subsubsection{Image and Context}

Some works combine image-based features with non-visual contextual features. \cite{dib2021multidimensional} adopts the method from \cite{nataraj2011malware} to convert binaries into grayscale images, while also extracting frequent byte-level tokens using NLP techniques. A CNN processes the images and an LSTM processes the token embeddings, each generating deep feature representations. These features are then fused into a joint feature vector, which is fed into a neural network, and the final classification result is produced by the output layer. FED-MAL \cite{abdel2023efficient} also belongs here: it embeds permissions, components, and API calls into semantic vectors by a bidirectional independently recurrent neural network (Bi-IndRNN), and transforms the vectors into grayscale images. The CNN-based spatial feature extraction is employed on the images to extract visual features. The semantic vectors and visual features are fused into a single capsule vector which is passed through a Sigmoid function for final detection.

\subsubsection{Graph}

\cite{petrache2025unveiling} evaluates graph-based similarity analysis for cross-platform malware classification (X86, ARM, MIPS). it first employs the Capstone disassembly engine to construct control-flow graphs are constructed from input binaries. On these CFGs, three types of embeddings are explored: node embeddings with Node2Vec and VERSE, edge embeddings obtained by applying the Hadamard product on pairs of node vectors, and graph embeddings generated by Graph2Vec. Results show that graph-level embeddings are more effective in capturing global structural patterns. Therefore, Once graph embeddings are obtained, malware classification is performed via similarity computation, including a proposed Greedy Algorithm for grouping highly similar samples.

\subsubsection{Summary}

\begin{figure}[htbp]
    \centering
    \includegraphics[width=0.75\columnwidth]{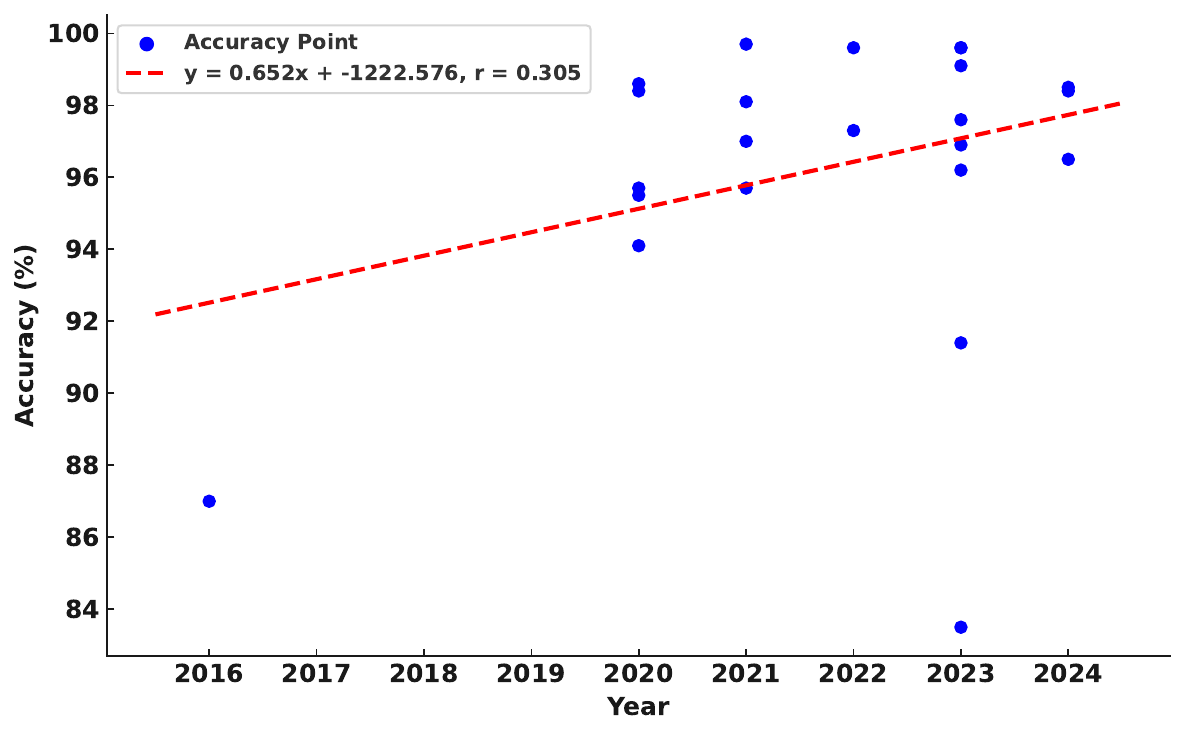}
    \caption{Year vs. Accuracy with OLS regression line for static analysis.}
    \label{fig:static_year_all_accuracy}
\end{figure}

\begin{figure}[htbp]
    \centering
    \includegraphics[width=0.75\columnwidth]{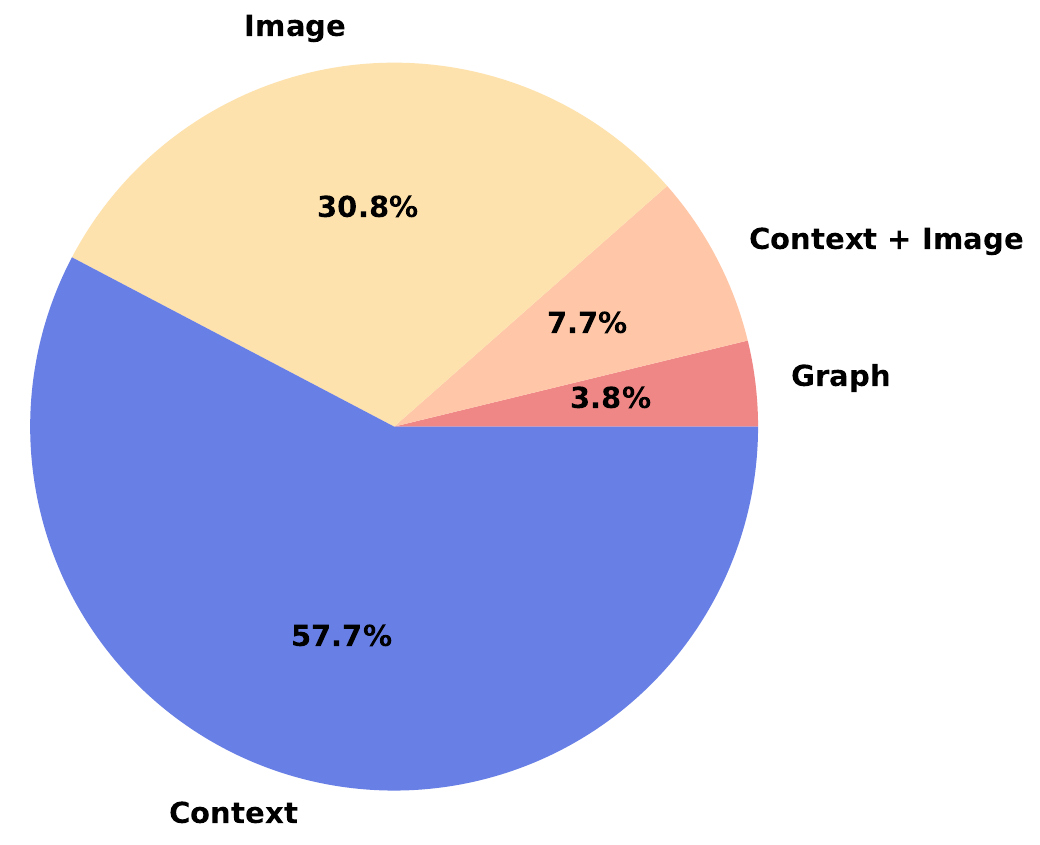}
    \caption{Distribution of feature types for static analysis.}
    \label{fig:static_feature_types_pie}
\end{figure}

In this section, we first summarized the general workflow of static analysis–based malware detection and classification, and then reviewed representative methods employing the three major categories of features: context-sensitive, image-based, and graph-based. We highlighted the similarities and differences in feature extraction techniques and analysis strategies across these works, thereby providing a comparative perspective on how different feature representations shape the effectiveness of IoT malware detection.
\begin{itemize}
    \item In the use of static features, context features and image features constitute the majority, accounting for 57.7\% and 30.8\%, respectively. A smaller portion (7.7\%) of studies adopt a hybrid approach that combines these two feature types, while only 3.8\% rely solely on graph-based features for malware classification. The distribution of these proportions is illustrated in Fig.~\ref{fig:static_feature_types_pie}.

    \item Different feature categories exhibit distinct strengths and limitations, and a single feature type is often insufficient to meet the requirements of IoT malware detection. Context-sensitive features can be extracted in a lightweight manner and are effective in capturing program semantics, but they are highly vulnerable to obfuscation, polymorphism, and platform heterogeneity. Image-based features alleviate these issues and are well suited for deep learning models such as CNNs and Vision Transformers, offering automated and scalable extraction. However, they fail to capture semantic information and entail relatively high computational and inference costs. Graph-based features not only preserve global semantic and structural properties of programs but also provide strong cross-platform generalization. Nevertheless, graph construction and vector embedding are computationally expensive, making them unsuitable for deployment on resource-constrained IoT devices. To overcome these limitations, multimodal fusion approaches that integrate semantic, visual, and structural information have emerged as a prevailing trend, offering a more robust and accurate solution.

    \item In IoT environments, context-sensitive features can be extracted and inferred locally on end devices, whereas image- and graph-based features generally incur higher overhead and thus require federated learning, edge deployment, or distributed computing frameworks for practical malware analysis.

    \item As illustrated in Fig.~\ref{fig:static_year_all_accuracy}, the accuracy of static analysis–based IoT malware detection and classification increased significantly between 2016 and 2020. However, after 2020, the improvement plateaued, with most reported results fluctuating within the range of 94\%–98\%. This trend is closely associated with the evolution of feature extraction techniques and the integration of deep learning methods. Prior to 2020, research efforts primarily relied on simple statistical representations and image-based approaches, which constrained the achievable accuracy. Starting in 2020, the introduction of deep learning architectures, such as n-gram–based CNNs and RNNs, along with semantic representations (e.g., BERT) and multimodal feature fusion, led to substantial accuracy gains. More recent studies have increasingly adopted federated learning, edge-based deployment, and other distributed frameworks to enhance robustness and scalability in IoT environments.
    
    \item In the future, the integration of context-sensitive, image-based, and graph-based features is expected to become a prevailing trend, addressing the inherent vulnerability of static features to obfuscation and packing techniques. Deep learning methods will play a crucial role in extracting deep semantic representations, while multimodal fusion combined with lightweight analysis is anticipated to enhance robustness and efficiency in IoT environments. This direction holds significant promise for advancing static analysis–based malware detection and classification.
\end{itemize}

\subsection{Dynamic analysis Features} 

\begin{figure*}[htbp]   
    \centering
    \includegraphics[width=1.0\linewidth, height=2.5cm]{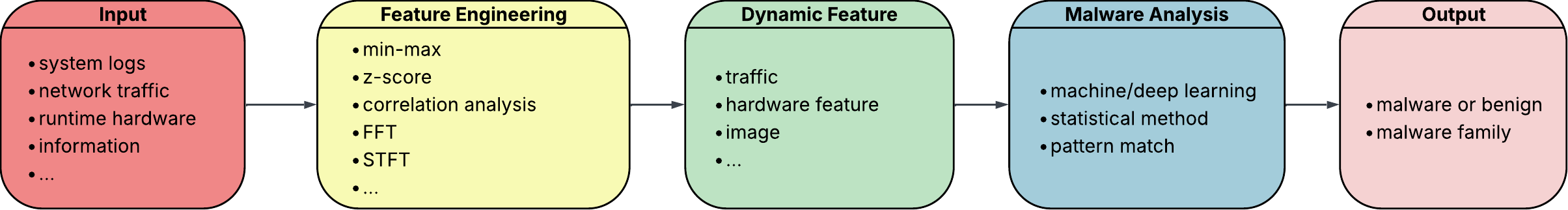}
    \caption{Workflow of dynamic analysis.}
    \label{fig:Dynamic_analysis}
\end{figure*}

Dynamic analysis has become a common and effective approach for malware detection and classification. A general workflow of dynamic analysis is shown Figure \ref{fig:Dynamic_analysis}.
Existing studies typically monitor running IoT devices to collect diverse sources of dynamic information, 
including \textit{system logs}, \textit{sensor data}, \textit{communication signals}, \textit{network traffic}, and \textit{runtime hardware information}. 
After feature engineering, these raw data are transformed into six major categories of features: hardware feature, runtime context, traffic,, image, spectral feature and graph.
After that, these features can either be directly utilized by traditional techniques 
such as pattern matching and statistical analysis, or embedded into vector representations and fed into 
machine learning or deep learning models for malware detection and classification.


\subsubsection{Hardware Feature}

Several studies \cite{kuruvila2021defending,demme2013feasibility,kadiyala2020lambda,elnaggar2022runtime,pan2022hardware} detect malware or anomalous behavior by collecting hardware performance counter (HPC) data or embedded trace buffer (ETB) outputs. Commonly collected metrics include Instructions Per Cycle (IPC), cache misses, and retired instructions, CPX and ETB signals. These raw traces are transformed into time-series vectors or matrices, and then classified with traditional ML or deep models. Different studies mainly differ in their feature engineering strategies. PREEMPT \cite{elnaggar2022runtime} employs SigSET \cite{6783318} to collect the lower 128 bits of the CPX as ETB signals. After cleaning the traces to remove resets and benign noise, the signals are converted into compact feature vectors, which are then classified using models such as KNN and Random Forest. In contrast, \cite{kuruvila2021defending} adopts a more sophisticated approach by applying three feature selection techniques—Univariate Selection, Feature Importance, and Correlation Matrix with Heat Map—to identify optimal subsets from the collected HPC data for malware detection.

While the above methods focus primarily on selecting or transforming ETB signals and HPC data, both \cite{demme2013feasibility} and \cite{pan2022hardware} emphasize temporal modeling of hardware traces. \cite{demme2013feasibility} transforms HPC-derived time series into feature vectors using four techniques: Raw Samples, Context-switch Aggregation, Histograms, and Scheduling-based Aggregation. Extending this idea, \cite{pan2022hardware} constructs a two-dimensional temporal matrix from combined HPC and ETB signals. In both cases, the resulting vectors or matrices are fed into classifiers such as Decision Trees (DT) to detect malicious behavior.

LAMBDA \cite{kadiyala2020lambda} further explores statistical characterization by executing predefined indicator programs (e.g., \texttt{ls}, \texttt{ps}, \texttt{netstat}) alongside the target and collecting HPC data with \texttt{perf stat}. The resulting matrix of CPU cycles, retired instructions, and related metrics is compared against a benign reference using Univariate and Multivariate T-Tests to detect malware.

Sentinel \cite{cosson2021sentinel} collects both system-level and process-level metrics via the \texttt{sysfs} interface. These metrics, including system-level information such as CPU frequency, memory usage, load averages as well as process-specific statistics such as physical and virtual memory usage, are stored in a PostgreSQL database. Sentinel continuously queries the PostgreSQL database and feeds the retrieved data into a trained model, such as Naive Bayes, PART(rule-based learning), multilayer perceptron (MLP), for malicious behavior detection.

Side-channel approaches \cite{soundar2018power, li2024zerod, sehatbakhsh2019emma} exploit physical-level leakages, such as power consumption or electromagnetic emissions, to infer malicious behaviors. Both \cite{soundar2018power} and ZeroD-fender \cite{li2024zerod} leverage power side-channel information for malware detection. \cite{soundar2018power} uses the Monsoon Power Monitor to collect power consumption traces from Android smartphones. Fast Independent Component Analysis (FastICA) is applied to extract independent signal components, and Recursive Feature Elimination (RFE) is employed to select the most informative features in each component. These features are then passed to a RF model which can distinguish benign and malicious traces. In contrast, ZeroD-fender extends this idea to IoT environments by combining static code analysis with power traces. It extracts shared behavioral features (e.g., shell commands and functions) from program code and collects corresponding power traces during execution. Then it applies a multi-window LSTM autoencoder to select representative traces, which are analyzed by a deep model (ThingNetV2) to detect zero-day malware. 

In contrast, EMMA \cite{sehatbakhsh2019emma} monitors electromagnetic (EM) signals during verifier–prover attestation. During proving process EMMA uses EM-Mon continuously to monitor the EM signals emitted by the Prover and extract features such as execution delay, EM spike frequency, and total attestation runtime. By comparing these features with a reference model, EMMA can flag deviations as potential attacks or integrity violations.

\subsubsection{Runtime Context}

Several approaches rely on direct feature vectorization. In \cite{Lee2021Precise, aboelwafa2020machine, sood2021accurate}, they directly convert sensor data into vectors for malicious behavior detection. PCoExtractor \cite{Lee2021Precise} encodes IoT device interactions into bit vectors (e.g., sensor states, value trends, actuator functions). Based on these vectors events are categorized into five predefined types, including Occurred, P-edge, N-edge, Uniform-1, and Uniform-0. anomalies are reported when event combinations or temporal transitions are absent from predefined valid correlation or transition databases. Similarly, in \cite{aboelwafa2020machine}, sensor readings collected during benign execution are directly generated into a vector which is used to train an autoencoder solely. If the reconstruction error for a new input vector exceeds a predefined threshold, it is classified as an false data injection (FDI) attack. Furthermore, \cite{sood2021accurate} analyzes data from a forest fire monitoring network, where meteorological readings with spatial coordinates are used to construct feature sets. Then it applies several binary classification algorithms such as Classification and Regression Trees (CART), Random Forest (RF), and Support Vector Machine (SVM) to distinguish normal from anomalous data. For samples classified as anomalous, Moran’s I spatial autocorrelation index is applied to further determine whether the anomaly is due to a sensor fault or an attack. 

In contrast, E-Spion \cite{mudgerikar2020edge} uses QEMU (Quick Emulator) to execute IoT programs and collects three types of behavioral logs, including the names of running processes, process-level behaviors such as execution time and resource usage, sequences of system calls made by processes. If the first type of features does not match the processes in the predefined whitelist, the latter two types of features are vectorized and fed into two Random Forest models respectively for anomaly detection. \cite{chulerttiyawong2023sybil} applies math calculation on collected communication signals in UAV (Unmanned Aerial Vehicle) network to derive Time Difference of Arrival (TDoA) and Received Signal Strength Difference (RSSD) features vectors from the UAV communications. The vectors are finally passed to multiple classifiers (J48, Classification via Regression, OneR, JRip) which differentiate genuine transmissions from Sybil attacks.

Other studies apply statistical and functional transformations to derive higher-level features from raw time-series or interaction data. Chakraborty \cite{chakraborty2021machine} segments SWaT water treatment sensor data into fixed time windows and then smooths them via spline fitting. Each smoothed segment is treated as a functional curve, and its first derivative (velocity) is computed. Functional Principal Component Analysis (FPCA) is performed on both the position and velocity curves to extract dynamic features such as level, slope, and peak values. These features are fed into multiple classifiers (M3–M6) for malicious data detection. Similarly, to preserve the temporal logic of events, AntiConcealer\cite{zhang2021anticoncealer} first aggregates, filters, encodes, and time-normalizes collected runtime data such as communication logs, SQL commands, and system logs to obtain feature vectors. It then applies a combination of Multivariate Hawkes Processes (MHP) and Bayesian Probabilistic Graphical Models (BPGM) for malicious attack detection. In contrast, Aegis+ \cite{sikder2021aegis+} targets smart home environments, where sensor–device interactions are engineered into trigger–action pairs and represented as discrete context arrays. A Markov Chain–based model is then trained on these arrays to learn baseline behavior patterns and detect malicious deviations.

Unlike traditional feature engineering approaches, \cite{jahromi2021toward} employs autoencoder-based techniques to extract features from sensor data. Specifically, two autoencoders are trained separately on normal and attack data. For an unseen instance, both models generate feature vectors that are concatenated into a joint representation, which is then reduced via PCA. The resulting compact vector is first evaluated by a One-Class SVM to detect anomalies, and subsequently classified as benign or malicious using a decision tree.

\subsubsection{Traffic}

A number of studies rely on pre-extracted features or preprocessed traffic provided by public datasets, or raw traffic captured in .pcap format using tools such as Wireshark, Tcpdump, or SnifferPro. These works typically perform lightweight preprocessing steps, including data cleaning, feature selection, normalization, and scaling, before feeding the data into learning models. For instance, both \cite{Wozniak2021recurrent, hasan2022securing, sedjelmaci2023secure, ahakonye2024classification} directly employs an existing dataset (e.g., N-BaIoT, UNSW-NB15 dataset, ICS-SCADA and CIRA-CIC-DoHBrw-2020) where all features are numerical. In \cite{Wozniak2021recurrent, hasan2022securing, sedjelmaci2023secure}, after applying data cleaning, vectorization and Min-Max scaling, the result feature vectors are fed into a neural network model for attack classification or detection. Notably, \cite{ahakonye2024classification} additionally applies SMOTE balancing to address class imbalance and employs PCC-based feature selection to remove redundant attributes. The resulting feature vectors are then fed into a CNN–LSTM model, which produces a three-class output, Non-DoH, Benign-DoH, or Malicious-DoH (including previously unseen zero-day attacks). Moreover, \cite{ahakonye2024trees} merges and cleans multiple datasets by using Python’s Pandas and NumPy libraries, then applies MinMaxScaler to normalize all feature values, which are fed into an explainable AI model for malicious traffic prediction. Furthermore, \cite{Guizani2020Network} removes incomplete entries and applies cross-correlation analysis to reduce features from 44 to 29 numerical ones, which are then used to train a three-layer RNN-LSTM to detect malicious traffic. 

To enhance representativeness, \cite{jiang2024blockchain} introduces a more comprehensive preprocessing pipeline that comprises data cleaning, domain knowledge-based feature engineering, sampling, label encoding, and Min-Max normalization, resulting in 91 protocol-level features that support federated learning with FedAvg. Along similar lines, both \cite{bhandari2022artificial} and \cite{mei2024novel} further applies Pearson Correlation Coefficient (PCC) analysis to select the 25 most informative features. The resulting vectors were then classified using models such as Support Vector Machines (SVM), Random Forests (RF), Deep Neural Networks (DNN), and even more complex architectures such as a seven-layer Multi-Layer Perceptron (MLP).

In contrast, another line of work emphasizes direct traffic parsing and feature vectorization from raw packets or flow records. For example, LEDEM\cite{ravi2020learning} extracts protocol types, flow counts, packet rates, and IP/port statistics, transforms them into vectorized tuples, and detects DDoS traffic using a Semi-supervised Stacked Deep Extreme Learning Machine (SDELM). similarly, MAFFIT\cite{ma2023multi} derives packet length and raw byte sequences, processes them via BiLSTM- and byte-sequence extractors, fuses the resulting vectors, and classifies traffic by cosine similarity against reference samples.  MalBoT-DRL\cite{alfawareh2024malbotdrl} applies Damped Incremental Statistics to obtain 23 features, normalized and fed into a Deep Q-Network for adaptive detection. Furthermore, after parsing each packet into a 7-tuple (e.g., direction, port type, packet length, protocol, inter-arrival time), DIOT\cite{nguyen2019diot} embeds it into discrete symbols based on device type using AuDI. As a result, the network behavior of an IoT device is transformed into a sequence of discrete symbols, which are subsequently fed into a type-specific GRU (Gated Recurrent Unit) model to determine whether the device's behavior is anomalous.

In contrast, \cite{vinayakumar2020visualized} extracts domain names from captured DNS packets, lowercases all characters and applies a character-to-index mapping, resulting in an indexed sequence. This indexed sequence is embedded into vectors using the Keras embedding layer and padded to a fixed length and then fed into a Siamese neural network, which filters out spoofed domain names. The final vectors are classified by an RNN/LSTM model to detect Domain Generation Algorithm (DGA)-generated domains which is commonly used by botnets.

Both IoT-Sentry\cite{malik2021iot} and \cite{ali2023effective} follow a similar pipeline of transforming raw traffic traces into structured features, normalizing them, and applying machine learning models for classification. IoT-Sentry converts PCAP traces into CSV format, extracts 21 cross-layer features, and applies quantile transformation combined with min–max scaling and finally feeds the normalized vectors into a Random Forest ensemble classifier.  WADAC\cite{sridharan2018wadac} uses scapy to extract packet- and traffic-level features such as Census, Ratio, Load, and Gap, trains an Autoencoder to detect anomalies based on reconstruction thresholds, and further applies Random Forest to classify attack types.

In contrast, \cite{ali2023effective} employs CICFlowMeter to extract flow-level attributes (e.g., source/destination IPs, ports, protocol, byte count) from pcap files, removes columns with high missing ratios, and groups the remaining attributes into three modalities: flow, flag, and packet features. After Min–Max normalization, recursive XGBoost and SULOV are used for feature selection, and the refined vectors are trained with a deep learning model to distinguish benign from malicious traffic.

Unlike other approaches that extract fine-grained features from raw traffic, BOTA\cite{Uhricek2023bota} operates at the flow level. It reconstructs bidirectional flows from large-scale exporters, converts them into vectorized representations, and applies a two-stage IDS where weak indicators flag suspicious attributes and a sandbox-trained meta-classifier integrates Boolean rules to identify compromised devices and issue interpretable alerts.

Richer pipelines incorporate dimensionality reduction, imputation, and federated training. For example, \cite{kaur2024intrusion} encodes IIoT traffic with one-hot and Min-Max scaling, trains GRU-based local models, and aggregates them via FedAvg for distributed intrusion detection. Both \cite{aljuhani2023deep} and \cite{mohamed2023digital} applies label encoding, and Min-Max scaling and feature selection to obtain final feature. The difference is that \cite{aljuhani2023deep} further compresses features with a Contractive Sparse Autoencoder, and detects attacks using BiLSTM, while \cite{mohamed2023digital} takes forensic images derived from structured traffic logs as input, and perform feature selection by via PCA with Extremely Randomized Trees. A CNN is then trained on the resulting vectors for final benign/malicious classification. 

Distinctively, \cite{alkadi2020deep} performs attack detection by analyzing alerts generated from deployed anomaly detection systems. Features such as source/destination IPs and ports are extracted from Host-based Intrusion Detection Systems (HIDS) and Network-based Intrusion Detection Systems (NIDS) alerts, normalized via min–max scaling into structured vectors, and subsequently processed by a BiLSTM model, which classifies each alert as either a malicious attack or abnormal network behavior.

Methods proposed in \cite{zhu2023lkd, lu2023two, abid2024multi} follow a similar workflow for attack classification. Traffic data are first cleaned and preprocessed through padding or truncation, then transformed via fixed-length encoding or Min–Max normalization to obtain feature vectors. These vectors are subsequently fed into deep learning models to identify attack categories including DDoS, backdoor, and C\&C traffic.

Several studies explicitly exploit temporal logic in feature construction by leveraging time windows, sequential patterns, or periodicity. For instance, \cite{hekmati2024correlation} builds traffic sequences which represents the number of packets sent within the $t$-th time window between two network nodes and augments them with neighboring node sequences to capture spatio–temporal correlations. The resulting traffic sequences fed into deployed models,such as MLP, CNN, LSTM, to produce a binary classification output indicating whether an attack is occurring in the current time window. Similarly, \cite{chang2023finish} organizes unsolicited Darknet traffic into matrices (rows = time, columns = IPs/ports) and applies federated Non-negative Matrix Factorization (NMF) to extract temporal patterns and spatial patterns, thereby enabling collaborative detection of coordinated scans. In \cite{moustafa2018ensemble}, it uses tcpdump to capture IoT traffic and extracts Protocol- and flow-level features—such as request-response pairs, the number of DNS queries, and TCP/IP metadata using Bro-IDS. A sliding window mechanism is then applied to construct sequential features with temporal context. The resulting feature set is fed into an AdaBoost ensemble with heterogeneous weak learners to identify malicious traffic instance. After capturing packets with Scapy and converting them into fixed-length time-series vectors through field masking and padding/normalization, ADRIoT \cite{li2021adriot} also uses sliding-window segmentation to construct sequential feature vectors, which are then analyzed with an LSTM Autoencoder using reconstruction error for anomaly detection.

Finally, \cite{niu2021Malware} focuses on DNS traffic by computing time-interval sequences of repeated domain requests. Similar intervals are normalized to identical values, and the resulting sequence is mapped into a character string representation. It finally detects periodicities via string matching and Fourier transform, thus capturing both explicit and hidden timing regularities for malicious domain identification.

Some studies calculate statistical features from the parsed traffic and use these features for attack or malicious traffic detection/classification. He et al.\cite{he2023lightweight} extract five-tuple and session-level statistics (e.g., duration, byte count, and flags), apply z-score normalization and aggregation. The final feature vector is fed into classifiers such as Decision Tree, Random Forest, KNN, and XGBoost to classify whether the traffic instance is attack-related or benign. Similarly, \cite{kundu2022detection} also uses the standard 5-tuple (source IP, source port, destination IP, destination port, and protocol) extracted from traffic stream and construct flows and sessions. It further computes various statistic features, such as  mean and standard deviation of packet and byte counts, session duration, and the final feature vector is fed into a trained one-dimensional Convolutional Neural Network (1D-CNN) to identify the presence of botnet-related activity and simultaneously classify the specific type of malicious behavior.  Passban\cite{eskandari2020passban} also follows the pipeline. It uses the NetMate tool to convert raw packets into flows and extract statistical features such as byte count, packet count, mean packet size, and flow duration, which are then evaluated using one-class models like Isolation Forest (iForest) and Local Outlier Factor (LOF). 

Several studies further process the extracted statistical features, either by performing additional computations or by applying feature selection techniques for dimensionality reduction \cite{sudheera2021adept,hafeez2020iot,shafiq2020corrauc}. Adept\cite{sudheera2021adept} profiles each device using session-level statistics such as the mean and standard deviation of flow sizes, detects anomalies via z-score analysis, and summarizes them into alert records. These alerts are further mined using Frequent Itemset Mining (FIM) to derive pattern-level features for attack classification. In \cite{hafeez2020iot}, after extracting 38 traffic statistics (e.g., packet and byte counts, unique IPs/ports, connection duration, and protocol type), IOT-KEEPER applies correlation analysis combined with the Correlation-based Feature Selection (CFS) method to remove redundancies, and employs Fuzzy C-Means (FCM) clustering to generate fuzzy rules for distinguishing benign and malicious traffic. CorrAUC\cite{shafiq2020corrauc} proposes a hybrid feature selection approach that evaluates correlations with class labels and redundancies among features, ranks them using an AUC-based wrapper method, and selects an optimal subset of representative statistical features using the Shannon Entropy TOPSIS method, which are then used by classifiers such as decision trees, Random Forest, or MLP for attack classification. FIOT\cite{han2024distributed} extends statistical feature extraction into the temporal domain by using a Kitsune-based extractor to derive 115 features over multiple time windows (100 ms to 1 min). These features are reduced via PCA and classified using a GRU-based neural network. 

FlowGuard \cite{jia2020flowguard} employs CICFlowMeter to extract 83 features per flow and selects the 10 most attack-relevant ones using information gain, which are compared against predefined filtration rules. Flows that match the rules are directly classified into the corresponding DDoS attack type, while unmatched flows undergo a traffic fluctuation–based anomaly detection algorithm that evaluates the deviation ratio between real-time and estimated stable traffic. Suspicious flows are then processed by an LSTM model to separate benign from malicious traffic; benign flows are forwarded, whereas malicious ones are further classified by a CNN into four DDoS categories.

Furthermore, ACID\cite{diallo2021adaptive} combines statistic features with semantic features to improve the robustness of classification and detection. It constructs bidirectional flows from raw packets and extracts statistical header features via  statistic computing, and optional semantic payload features via Word2Vec and TextCNN. These features are embedded by an adaptive clustering module that learns low-dimensional embeddings with supervised and contrastive objectives. The enriched representations are fed into a RF to detect and categorize malicious traffic.


\cite{cai2023adam, siniosoglou2021unified, mitev2020leakypick, yin2019connspoiler} further process statistics derived from traffic data to obtain features that better capture network behavior. Both \cite{cai2023adam, siniosoglou2021unified} start by capturing network traffic using tools such as Tshark, but their feature extraction and detection strategies differ. \cite{cai2023adam} derives seven header-level features (e.g., IP, protocol, port), computes entropy vectors (EVs), and applies unsupervised methods (KNN, K-means, Isolation Forest) to detect deviations from a nominal baseline. In contrast, \cite{siniosoglou2021unified} combines time-series data retrieved from SCADA systems via REST APIs with flow-level statistical features extracted by CICFlowMeter to form multimodal feature vectors, which are then evaluated using an Autoencoder-GAN and further classified with a Conditional GAN (cGAN) into specific attack types. 

\cite{yin2019connspoiler, mitev2020leakypick} follow a similar principle of leveraging statistical and distributional traffic features, but they are applied in different environments. ConnSpoiler \cite{yin2019connspoiler} monitors DNS traffic from devices and records all DNS queries that return NXDOMAIN (non-existent domain) responses. It then performs character-level analysis, calculating the likelihood of the domain being malicious or benign based on character frequency and bigram transition probabilities. If the score exceeds an upper threshold, the domain is classified as malicious(botnet-infected). LeakyPick \cite{mitev2020leakypick} collects network traffic before and after audio playback (including packet size, direction, inter-arrival time, and address information), and then detects whether smart home devices are covertly uploading audio data by either comparing traffic bursts before and after playback or applying a two-sample t-test to analyze statistical features such as packet size and inter-arrival time for significant distributional changes.

C2Miner \cite{davanian2024Cminer} is a sandbox-based framework for executing IoT malware and identifying Command and Control (C2) servers. The system intercepts outbound connections via a Man-in-the-Middle module, redirecting IP- and DNS-based requests to a designated IP:port space. A profiler then filters irrelevant traffic, records access frequency, and assigns each destination a C2 likelihood score. To confirm genuine C2 servers, C2Miner employs two strategies: a SYN-DATA-aware analysis of handshake and data exchange patterns, and a fingerprinting-based approach that derives communication grammars. The framework outputs active C2 server addresses along with fingerprint tags that aid network forensics and attribution.


Both \cite{meidan2022cadesh} and \cite{dai2023cmftc} share the philosophy of leveraging multi-dimensional features and constructing flow-level representations for malicious traffic analysis. In \cite{meidan2022cadesh}, traffic is first represented using the IP Flow Information Export (IPFIX) format and enriched with multiple feature types, including raw attributes (protocol type, packet count, flow duration), derived statistics (e.g., inter-arrival time), DNS-based characteristics, reputation intelligence, and temporal collaborative features. These multi-source features are first fed into an autoencoder, if the mean squared error (MSE) of a new input exceeds a predefined threshold, the traffic is considered infrequent and is passed to a pre-trained K-Means clustering model for malicious traffic identification. In \cite{dai2023cmftc}, encrypted traffic is characterized by two complementary modalities: payload content and protocol headers. Payload features are extracted through bidirectional flow-relative position encoding and a Cross Block module, producing both semantic vectors and high-dimensional representations. Header features (e.g., payload length, TCP window size, inter-packet intervals, directionality) are concatenated with the semantic vectors, then processed using depthwise 1D convolutions to capture temporal dependencies. These enriched feature sequences are integrated with payload representations to form global flow-level semantics, which are finally leveraged in a multi-task attack classification framework.

Unlike traditional approaches, \cite{aouedi2022federated, yuan2023mcre} focus on extracting low-dimensional representations from raw features for subsequent analysis. Specifically, \cite{aouedi2022federated} employs a federated learning framework and trains an autoencoder optimized by minimizing reconstruction error. This autoencoder learns compact representations from 23/26-dimensional industrial control system features (e.g., pressure, valve state, flow rate, sensor readings) in a distributed setting. The resulting latent vectors are then fed into a fully connected network (FCN) for binary classification of benign versus malicious activity. In contrast, \cite{yuan2023mcre} addresses noisy-labeled traffic data using a 7-layer fully connected encoder to derive low-dimensional embeddings, further refined with three constraints: minimizing reconstruction error through an autoencoder structure, maximizing the normalized inter-class margin relative to intra-class distance, and enforcing intra-class compactness around high-confidence samples. The learned vectors are finally clustered with K-Means, enabling robust categorization of malicious traffic.

FS3 \cite{ayesha2023fs3} employs a supervised self-supervised learning strategy to derive latent vector representations. It first transforms raw network traffic into a tabular format that combines numerical features (e.g., packet size) and categorical features (e.g., protocol type). A pre-trained TabNet model is then used to capture feature dependencies within the table and map each sample into a compact latent representation. Finally, these latent vectors are classified using a subsampled k-Nearest Neighbor (KNN) algorithm to distinguish benign from malicious traffic.

Unlike the above-mentioned work, \cite{shen2018multistage} proposes a three-layer intrusion detection system that incorporates a privacy-aware, game-theoretic detection mechanism. The system first collects audit data (i.e., network behavior, including traffic logs) at the IoT layer. This audit data is then uploaded to the fog layer, where a signaling game named MPMDG is used to decide whether to preserve user privacy (by modifying the audit data before upload) or to directly forward the unaltered data to the cloud layer for intrusion detection. Finally, the cloud layer employs a deployed IDSaaS (Intrusion Detection System as a Service) to analyze the data received from the fog layer and determine whether malicious activity is present. This work focuses on investigating how the false positive rate at the fog layer and the detection rate at the cloud layer affect the system's ability to defend against the spread of malicious behavior.

\subsubsection{Spectral Features}

Several works \cite{sehatbakhsh2020remote,chawla2021machine,nazari2017eddie,pham2021obfuscation} leverage electromagnetic (EM) side-channel emissions, converting raw signals into spectral representations(SRs) using Fourier transforms (FFT or STFT). In the malware analysis stage, these spectral representations are processed using different techniques. In \cite{sehatbakhsh2020remote}, spectral representations are clustered with HDBSCAN and the CAPE distance metric and if any new representations derived from incoming EM signals is considered anomalous if more than half of its samples do not match any of the previously clustered regions. similarly, \cite{nazari2017eddie} directly applies statistical testing via the Kolmogorov–Smirnov (K-S) test on the obtained SRs for malware detection. In contrast, \cite{pham2021obfuscation} and \cite{chawla2021machine} applies Normalized Inter-Class Variance (NICV)-based feature selection and PCA on the obtained SRs to reduce dimensionality and select optimal SRs respectively. Then \cite{pham2021obfuscation} fed these SRs in MLP and CNN models, while \cite{chawla2021machine} applies Random Forests and SVMs on these SRs for malware type classification.

In \cite{bai2022rascv2}, STFT is also used to convert power consumption signals which is collected via an onboard Analog-to-Digital Converter (ADC) into time–frequency spectrograms. Spectrogram vectors are inputted into a supervised Support Vector Machine (SVM) classifier to distinguish between benign execution and different categories of injected attacks. \cite{vijayakanthan2023fortifying} also extends this idea. It first converts runtime memory snapshots into audio waveforms. The resulting signals are processed with STFT and additional spectral feature extraction methods such as MFCCs, Mel spectrograms, and Chroma representations, which are subsequently classified by a pre-trained CNN to identify malicious content.

In contrast to STFT-based feature engineering, pAElla \cite{libri2020paella} continuously samples current and voltage via an ADC, computes the Power Spectral Density (PSD) using the Welch method, and performs lightweight on-device malware detection through ARM’s NEON Single Instruction Multiple Data (SIMD) extensions. DeepPower \cite{ding2020deeppower}, by contrast, refines raw power traces using a robust autoencoder and filtering methods(a 100-point Simple Moving Average (SMA) filter and a wavelet denoising method), converts them into mel-scaled spectrograms, and applies CNN and LSTM models to infer device behaviors, which are then compared against a predefined infection model to determine compromise.

\subsubsection{Image}

Existing image-based approaches for malicious traffic analysis generally follow a similar pipeline. Raw traffic captured in .pcap format is first transformed into fixed-length byte sequences, which are then converted into images by using tools like USTC toolset or IDX representations. Among these, some studies adopt grayscale images \cite{zhang2024enhanced,zhang2024automatic,ning2022malware,huo2024lightguard,xu2024self}, while others employ RGB images \cite{hu2023Deep}. Deep neural networks are subsequently leveraged to extract semantic representations, and the final classification is performed using either a softmax layer or a separate classifier such as logistic regression (LR).

More specifically, in the traffic-to-image transformation stage, studies such as \cite{zhang2024enhanced,zhang2024automatic,ning2022malware,huo2024lightguard,xu2024self} generally adopt a multi-step preprocessing pipeline that includes traffic segmentation, cleaning, image generation, and IDX file construction, although the details vary slightly across works (e.g., session truncation in \cite{huo2024lightguard}, or application-layer slicing in \cite{ning2022malware}). In the feature learning stage, most methods employ convolutional neural networks (CNNs) to capture deep spatial and semantic patterns from traffic images. Some studies further enhance CNN-based representations with lightweight architectures \cite{huo2024lightguard}, neural architecture search \cite{zhang2024automatic}, or semi-supervised/domain-adaptive components \cite{ning2022malware}. In particular, Ning’s framework \cite{ning2022malware} augments CNNs with a Ladder Network to exploit both labeled and unlabeled data, incorporates feature-level domain adaptation for cross-domain robustness, and extends shallow CNNs to deeper VGG-based variants (KTDA-VGG and KTDA-VGGLadderNet), which significantly improve classification performance. Notably, \cite{xu2024self} further employs a Vision Transformer (ViT) encoder to obtain a classification token from IDX files, which serves as the final feature vector. Finally, the learned feature vectors are fed into either a softmax classifier or, in the case of \cite{zhang2024enhanced}, a logistic regression (LR) model to produce the final traffic category.

In \cite{hu2023Deep}, malicious traffic packets are similarly processed into images but represented in RGB space using the USTC toolset. These images are then input into the DSAN-AT model, which integrates Local Maximum Mean Discrepancy (LMMD) for unseen malicious traffic and a Convolutional Block Attention Mechanism (CBAM) to highlight informative regions. Finally, DSAN-AT uses a softmax function to predict the probability distribution over traffic classes.

Several studies apply different methods to transform side-channel information into image representations for malware analysis. For example, LightAuditor \cite{jung2022light} processes the collected power traces through three steps: (1) applying a Root Mean Square (RMS) method to remove alternating current (AC) fluctuations, (2) using an alpha filter with a smoothing factor of 0.2 to suppress high-frequency noise, and (3) converting the filtered traces into Continuous Wavelet Transform (CWT) images. However OD-meth \cite{albasir2023toward} first transforms the collected one-dimensional power signals into two-dimensional time-frequency representations (TFRs) using Constant Q Spectral Transformation (CQT), and then applies Histograms of Oriented Gradients (HOG) to extract high-level structured features, resulting in HOG feature images. Unlike these power-oriented approaches, \cite{kasarapu2023resource} leverages hardware performance counters (HPCs), applies PCA for feature selection, and directly maps the selected values into grayscale images. In all cases, the resulting images are analyzed by CNNs for final malware classification.

unlike aforementioned works, SQA \cite{esmaeili2022iiot} targets detecting black-box adversarial attacks in IIoT systems by converting raw queries into bytecode and then directly maps bytecode to grayscale images, enabling a unified input format for neural processing. A CNN-based encoder trained with contrastive loss (Mahalanobis distance) maps these images into semantic embeddings, which are compared with historical queries via k-NN matching. Queries deemed similar to malicious samples are blocked, while others are further classified by a CNN to confirm benign or malicious behavior.

\subsubsection{Graph}

A graph is a higher-level abstraction that can capture not only the network interactions among devices in an IoT environment but also the data exchanges at the system level. Communication-oriented frameworks, ContraMTD \cite{han2024contramtd}, GODIT \cite{paudel2019detecting}, and Kalis2.0 \cite{rullo2023kalis2}, all start from traffic collection and progressively transform raw packets into graph-based representations, but differ in their feature processing and detection strategies. ContraMTD aggregates flows into sessions and channels, models the network as a host interaction multigraph, and extracts local temporal channel features from the channels (via Fully Convolutional Neural Network (FCNN)) and global features from the multigraph (via Graph Double Edge Attention Network(DE-GAT)). ContraMTD leverages contrastive learning to enforce consistency between local and global features, flagging traffic as malicious if this consistency is violated. GODIT instead converts pcap traffic into communication graphs for DoS attack detection. It applies random walks and node2vec to obtain path embeddings, and uses Shannon entropy to select discriminative features before anomaly detection with a Robust Random Cut Forest (RRCF) model. Kalis2.0 extends this idea by collecting both traffic and contextual attributes such as port scans, banner grabbing, penetration testing) to build device profiles. These profiles are combined in the cloud to construct a Directed Acyclic Attack Graph (DAAG), whose structure provides high-level graph embeddings for context-aware analysis. Detection strategies are then deployed at the edge, where a hybrid engine combining rule-based and anomaly-based modules identifies Telnet brute-force attempts, abnormal traffic, and suspicious control connections. In the routing domain, Tong \cite{tong2023novel} extracts traffic features in RPL-based IoT networks to construct Destination Oriented Directed Acyclic Graphs (DODAGs), where nodes maintain baselines (Time-To-Live (TTL), hop count, latency, arrival rate) and collaboratively detect and recover from wormhole attacks through rank-based analysis and graph reconstruction. 

Beyond traffic, Wuchner \cite{wuchner2014malware} monitors system calls to build Quantitative Data Flow Graphs (QDFGs) which includes four types of system entities (i.e., Process, File, Socket, and Registry) as nodes and data flows between entities as edges. It applies three types of heuristic rules including,replication heuristics, manipulation heuristics, and quantitative heuristics, to determine whether the observed system behavior matches known malicious patterns.

\subsubsection{Summary}

\begin{figure}[htbp]
    \centering
    \includegraphics[width=0.95\columnwidth]{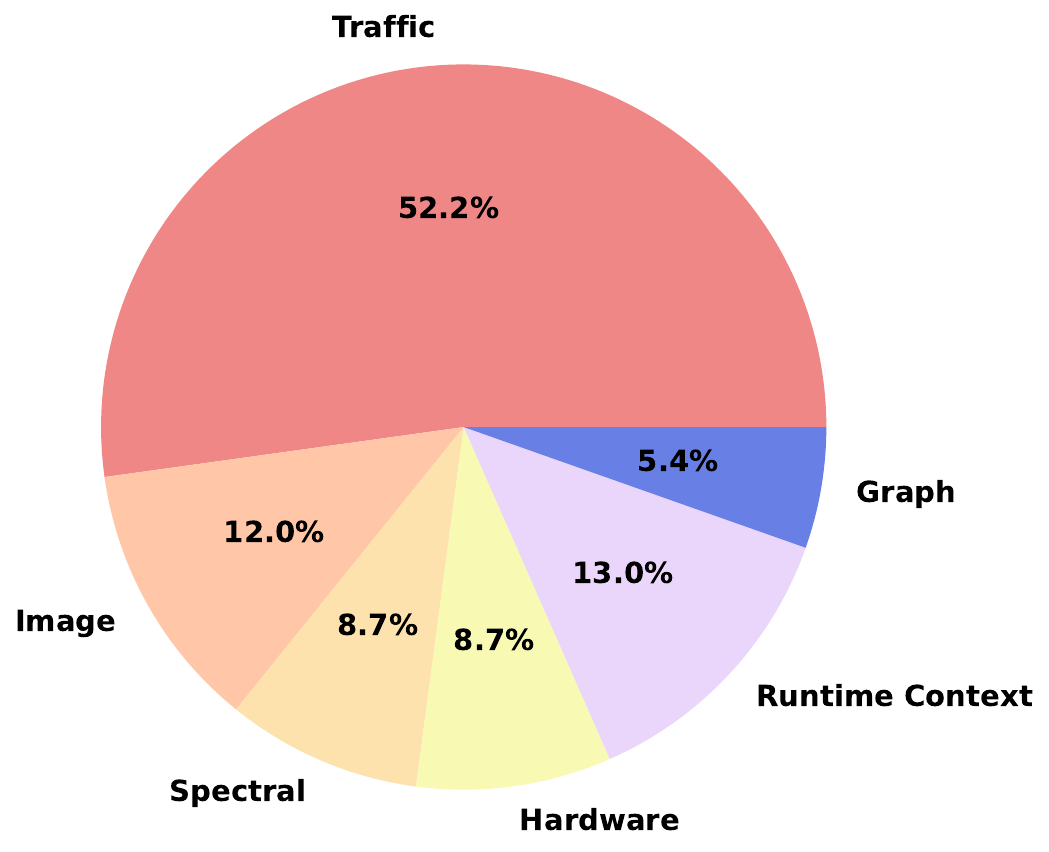}
    \caption{Distribution of feature types for dynamic analysis.}
    \label{fig:dynamic_feature_types_pie}
\end{figure}

\begin{figure}[htbp]
    \centering
    \includegraphics[width=0.75\columnwidth]{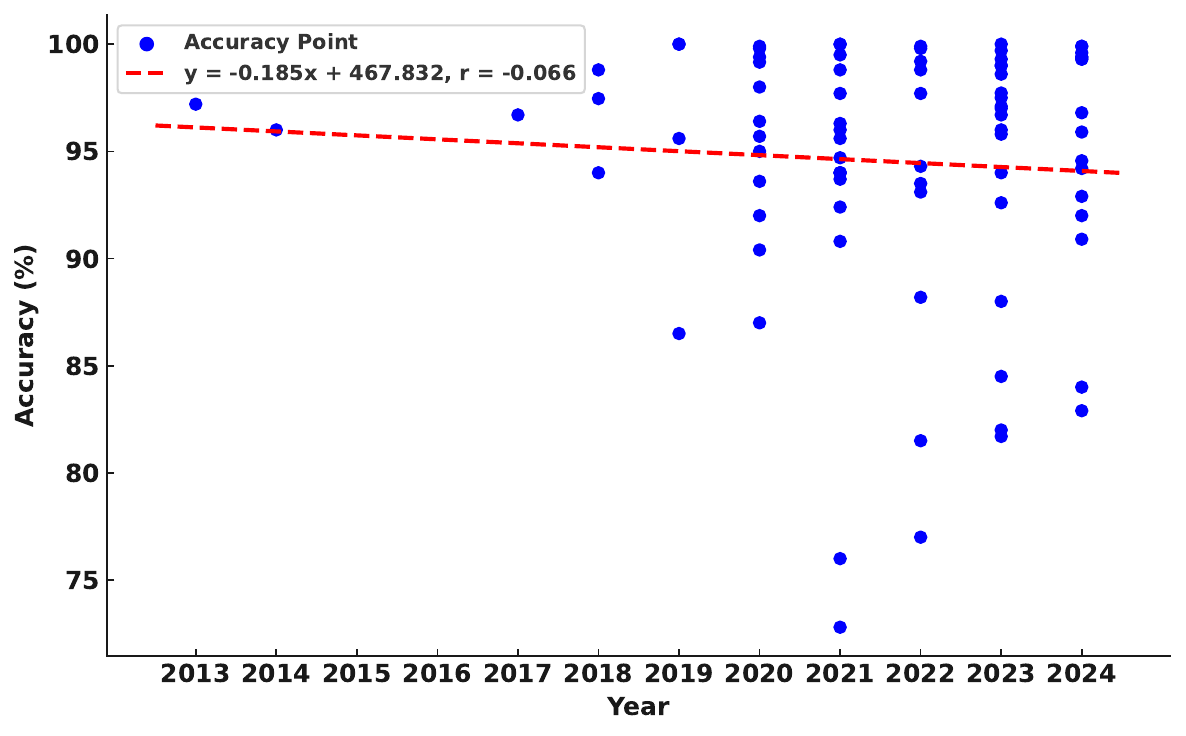}
    \caption{Year vs. Accuracy with OLS regression line for dynamic analysis.}
    \label{fig:dynamic_accuracy_regression}
\end{figure}

In this section, we first outline the workflow of dynamic analysis and then categorize existing methods into six feature families: hardware features, runtime context, traffic, spectral features, image-based representations, and graph-based representations. We further highlight the similarities and differences in feature extraction techniques and malware analysis strategies across these categories.
\begin{itemize}

    \item As illustrated in Fig.~\ref{fig:dynamic_feature_types_pie}, among the reviewed dynamic-analysis studies, traffic-based approaches dominate, accounting for 52.2\%. Image-based and runtime-context methods follow at 12.0\% and 13.0\%, respectively. In contrast, spectral, hardware-based, and graph-based features are used less frequently, representing 8.7\%, 8.7\%, and 5.4\%, respectively.

    \item Using different feature types for IoT malware detection has distinct benefits and limitations. For example, most studies detect attacks such as DoS/DDoS and botnets using traffic features, largely because traffic is relatively easy to collect, supported by a mature tooling ecosystem (e.g., Wireshark, tcpdump, Zeek, CICFlowMeter), independent of the instruction set, and thus offers strong cross-platform generalization. Moreover, given the characteristics of IoT deployments, network-centric attacks like DoS are very common, which further drives the widespread use of traffic features in this area. However, with the growing adoption of encryption protocols (e.g., TLS, QUIC, DoH), the semantic content of traffic becomes harder to extract, leaving mainly metadata and timing patterns. In addition, traffic primarily reflects inter-device behavior and provides limited visibility into the program-level behavior within individual devices.

    \item Other feature families also present distinct strengths and limitations. Hardware and spectral features capture micro-architectural and physical side-channel signals (e.g., power, EM, acoustic), revealing device behavioral logic and complementing traffic’s blind spots. However, device-permission requirements and firmware heterogeneity make these signals difficult to acquire and limit their applicability. Runtime-context features are not constrained by such access issues; they characterize trigger–action chains and system behavior with high interpretability, making them suitable for on-device modeling in IoT. That said, they are susceptible to environmental noise and operational variance. For image- and graph-based features, extraction and processing pipelines are relatively mature, but practical obstacles remain: image encodings preserve semantics of raw traffic and sequences yet struggle to capture temporal logic, which limits their utility for attack detection; graph construction faces an inherent fidelity–resource trade-off—higher precision demands substantial compute and memory, often untenable on resource-constrained IoT devices.

    \item We also analyze how detection/classification accuracy has evolved over time, as shown in Fig.~\ref{fig:dynamic_accuracy_regression}. We observe that the performance of dynamic-analysis techniques has remained largely stable, without significant improvements attributable to advances such as deep learning.

    \item In summary, traffic features have been extensively studied in terms of extraction and processing techniques, while visualization-based representations of hardware and runtime context features remain an emerging research direction. However, two critical challenges persist: (1) hardware information often contains substantial noise, and developing effective noise reduction techniques while preserving discriminative characteristics remains difficult; and (2) the collection of hardware and context features typically requires deploying specialized programs or devices on the target IoT system, which introduces resource overhead. Reducing such overhead while maintaining detection accuracy is an important challenge for future research.

\end{itemize}

\subsection{Hybrid Analysis Feature}

\begin{figure*}[htbp]   
    \centering
    \includegraphics[width=0.9\linewidth, height=4cm]{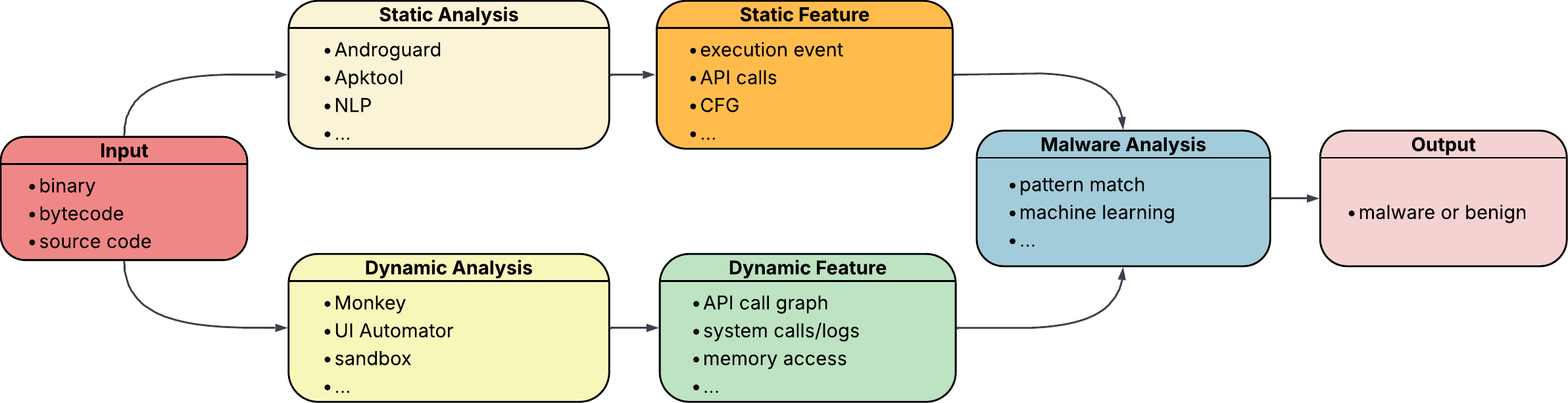}
    \caption{Workflow of hybrid analysis.}
    \label{fig:Hybrid_analysis}
\end{figure*}

In existing research, many studies rely on a combined use of static and dynamic analysis to detect malware in IoT environments. A general workflow of hybrid analysis is shown in Figure \ref{fig:Hybrid_analysis}. Typically, static analysis is first applied to parse raw binaries, source code, or bytecode, extracting contextual information such as API calls, abstract syntax trees (ASTs), or control flow graphs (CFGs). In some approaches, these static artifacts are directly compared with those of benign programs using techniques such as pattern matching to identify potential malware. Once suspicious behavior is detected, the program may then be executed and monitored at runtime to improve detection accuracy. Other approaches instead use the static information to guide dynamic execution within controlled environments such as sandboxes or Genymotion emulators. Dynamic behaviors are then captured through runtime monitoring or binary instrumentation, and together with static features, they are analyzed using methods such as temporal analysis or state transition analysis. Ultimately, static and dynamic features may also be fused into a hybrid representation, which is then fed into a trained machine learning–based model for malware detection.


\subsubsection{Context-sensitive Feature}

Existing studies adopt different strategies to combine static and dynamic program information for malicious behavior detection. Some approaches rely on static analysis (e.g., using tools such as AstBuilder) to parse source code or APK files and extract the expected behavior patterns of benign programs. During detection, the program is dynamically executed and the observed runtime behavior is compared against the pre-extracted patterns to identify potential malicious actions \cite{sikder2019aegis, lu2017time, zhang2018homonit}. Other studies directly integrate features obtained from both static and dynamic analyses to train a unified model for inferring malicious behaviors \cite{wazid2022bsfr, 8930038, iqbal2022rthreatdroid, ding2021mini}. More advanced methods perform a two-step process: first, static features are used for an initial classification to determine whether a program is suspicious, and then dynamic features are employed either to further evaluate the severity of the maliciousness \cite{deng2023trusted} or to re-verify the classification decision for improved accuracy \cite{9976924}.

For each SmartApp under inspection, Aegis\cite{sikder2019aegis} performs static analysis of its source code to extract the trigger-action logic, which is formalized as the App Context. The collected runtime data and the App Context are then combined to construct a context array, representing the runtime correlations among program states, sensors, and devices. Each context array is processed by a Markov Chain model to compute state transition probabilities, transitions with extremely low probability are flagged as anomalous or malicious behaviors. 
HoMonit\cite{zhang2018homonit} first derives the expected deterministic finite automatons (DFAs) of benign SmartApps. For open-source SmartApps, the AstBuilder tool is used to convert the source code into an abstract syntax tree (AST), which is then translated into a DFA. For closed-source SmartApps, natural language processing (NLP) techniques are applied to parse the textual description into a DFA. These DFAs capture the expected program behavior, including initial states and valid state transitions during execution. At runtime, device communications are captured and converted into event sequences, which are then checked against the expected DFAs. Any deviations, such as unauthorized commands or invalid state transitions, are flagged as attacks.

In contrast, \cite{lu2017time} integrates system requirements with program binaries to build a Runtime Security Model (RSM). Execution events are categorized into branch, trigger, regular and pair, and further semantically classified into events, control flow events, nullified control flow events, and mimicry malware sequences. From requirements, Best-Case and Worst-Case Execution Times (BCET, WCET), event order, and temporal relations are extracted to construct the RSM. To enhance efficiency, three greedy algorithms select representative RSM* subsets based on time range, variance, and peak probability, which are deployed in hardware for monitoring. During execution, the RAD hardware collects trace-port data and performs sequence, timing, and nullified-event checks to detect malicious behavior.

In program behavior representation, combining dynamic and static features enables a more comprehensive abstraction. In \cite{wazid2022bsfr}, dynamic features (e.g., system logs, execution traces) and static features (e.g., API calls) are integrated to train the detection model. A honeypot collects data from ransomware and benign programs, which, after cleaning, are transformed into two feature sets: signature-based features (SigRW), such as hashes and rules, and static vector features (FTRW), such as code structures and API calls. These features are used to train traditional machine learning models, including Random Forest, Logistic Regression, and Decision Tree. At detection time, the honeypot monitors runtime API calls, extracts corresponding feature vectors, and applies the trained models to classify samples as ransomware or benign.

In \cite{8930038} and \cite{iqbal2022rthreatdroid}, the authors leverage both static and dynamic analyses to characterize Android applications. Specifically, \cite{8930038} employs Androguard to decompile APK files and extract static features such as permissions and API calls, whereas \cite{iqbal2022rthreatdroid} uses APKTool to decompile APKs and extract the AndroidManifest.xml, .dex files, and resource files. From these artifacts, regular expression matching is applied to retrieve API calls, permissions (e.g., INTERNET, BIND\_DEVICE\_ADMIN, READ\_PHONE\_STATE), and embedded threat-related text, which are then used to construct static feature vectors. To capture dynamic behaviors, both studies execute the applications in a Genymotion emulator, where the monkey tool is used to generate random events and stimulate program execution. A key distinction is that \cite{iqbal2022rthreatdroid} additionally captures runtime screenshots and applies OCR techniques to identify threatening messages or diamlogs; the extracted text is then combined with text from resource files to build dynamic feature vectors. After obtaining both static and dynamic features, these vectors can either be directly concatenated into a unified representation and fed into classifiers such as Support Vector Machine (SVM), Random Forest (RF), or Naïve Bayes (NB), or further preprocessed through steps such as text tokenization, noise removal, and image normalization (e.g., scaling, grayscaling) to ensure compatibility with classification models. Finally, the processed features are input into pre-trained classifiers, including SVM, Decision Tree, RF, NB, and Logistic Regression, to determine whether the application is ransomware or benign software.

Furthermore, Mini-Me \cite{ding2021mini} addresses data-oriented attacks on robotic aerial vehicle (RAV) control programs through three stages: instrumentation, model construction, and online detection. The binary is lifted to LLVM IR to build a CFG, where symbolic taint analysis identifies target mathematical functions. Dynamic instrumentation (e.g., Valgrind) records call frequencies, and static data-flow analysis extracts critical state variables, defining input/output sets for LSTM-based modeling. Normal execution traces are collected, normalized, and segmented with a sliding window to train a lightweight LSTM that learns input–output mappings, optimized with mean squared error and Adam, and later quantized for deployment. At runtime, monitoring functions predict outputs from real-time inputs. Any deviations beyond a benign confidence interval indicate compromise, triggering fail-safe mechanisms.

Some approaches leverage dynamic analysis to validate and refine static analysis results. ITEC \cite{deng2023trusted} first extracts runtime signatures including comprising system calls (e.g., NtCreateProcess, NtCreateThread, NtListenPort), memory access patterns, and network behaviors, and then compares them against a database of known malware and benign signatures. Suspicious samples are executed in a sandbox, where a threat score is computed across exploitability, impact, and environment dimensions to classify risk levels (low, medium, high).

Similarly, in \cite{9976924}, the APK is decompiled with Apktool to obtain AndroidManifest.xml and .dex. The Androidguard extractor derives static features including API requests and network indicators (IPs, emails, URLs) from .dex, and intents/permissions from the manifest, then writes them to text and converts the text into a fixed-dimensional numeric vector for a pre-trained classifier. If predicted benign, the app undergoes dynamic analysis in an Android emulator, where Monkey/UIAutomator drives interactions. An API call graph is then built (node means system calls with frequency attributes and directed edge shows execution order), and four heterogeneous sub-models (Permission, Intent, API, Hardware) process their respective feature views. Finally the outputs are fused to yield the final classification.

\subsubsection{Summary}\leavevmode\\

\begin{figure}[htbp]
    \centering
    \includegraphics[width=0.45\textwidth]{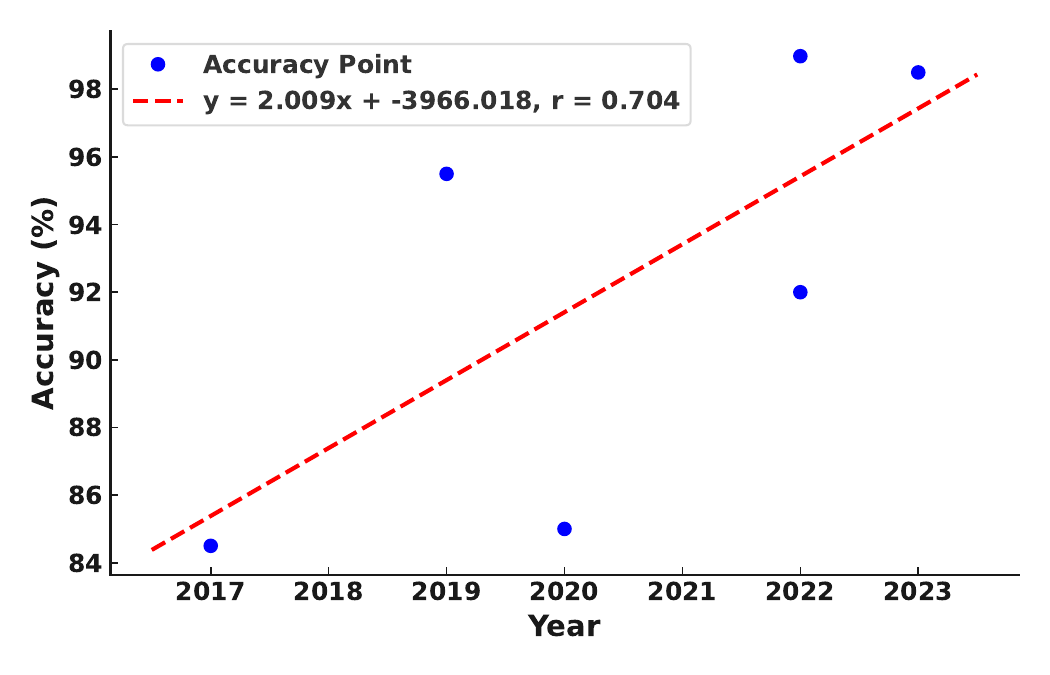}
    \caption{Year vs. Accuracy with OLS regression line for Hybrid analysis.}
    \label{fig:context_sensitive_regression}
\end{figure}

In this section, we first described the workflow of hybrid analysis and then discussed the methods that employ context-sensitive features, highlighting both the similarities and differences in feature extraction techniques and malware analysis approaches across the reviewed studies.
\begin{itemize}
    
    \item Existing hybrid analysis methods for malware detection are all built on context-sensitive features. They mainly adopt three hybrid strategies. The first extracts expected behaviors of benign programs through static analysis (e.g., trigger–action logic, DFA), and then compares them with the observed runtime behaviors of the target program to detect deviations that may indicate malware. The second integrates static and dynamic features into a unified representation, which is then fed into ML/DL models for classification. The third employs a two-stage process where static analysis performs an initial coarse screening, followed by dynamic analysis to refine or validate the detection results.

    \item In Figure~\ref{fig:context_sensitive_regression}, we report the studies that employed accuracy as the evaluation metric and observe that accuracy has consistently improved over time. This trend can be attributed to three main factors. First, feature engineering techniques have become increasingly comprehensive. Early approaches relied on relatively simple designs, primarily extracting static rules through customized algorithms, such as constructing event state transition models based on temporal logic. Subsequently, researchers explored statistical methods and traditional machine learning techniques, leveraging Markov Chains to model probabilistic state transitions and employing classifiers to capture more complex behavioral patterns, thereby enhancing detection accuracy. More recently, deep learning and multimodal feature fusion techniques have been introduced to extract richer and more representative program behaviors. For example, recurrent neural networks such as LSTM have been utilized to model input–output relationships, while multimodal features combining text, images, and APIs have been integrated for classification. These advances have further contributed to the improvement of malware detection accuracy.
    
    \item Existing studies mainly leverage these hybrid features for malware detection, while their application to malware classification has received little attention. Future research could explore alternative combinations, such as integrating image-based features with dynamic contextual information, to enhance both malware detection and classification.
\end{itemize}

\subsection{Graph Learning Representation Features}

\begin{figure*}[htbp]   
    \centering
    \includegraphics[width=1\linewidth, height=4cm]{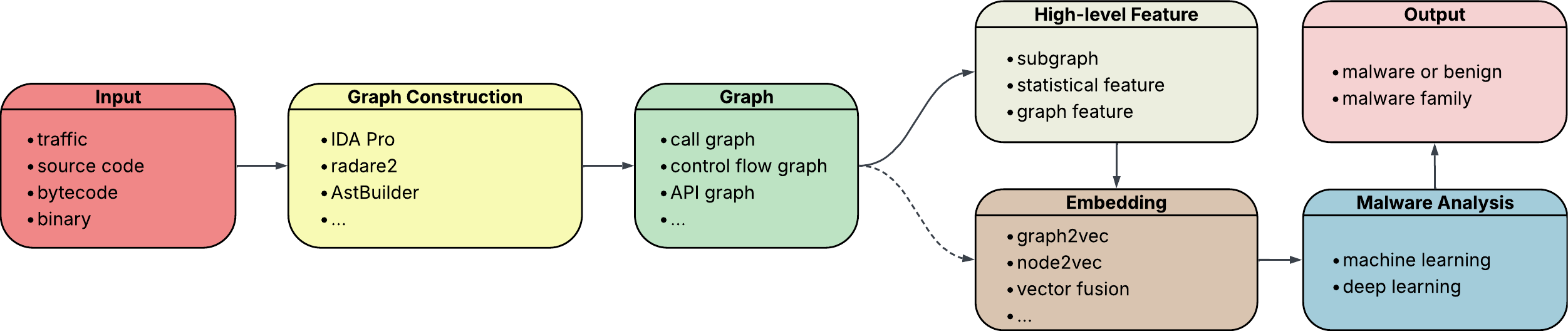}
    \caption{Workflow of Graph learning method.}
    \label{fig:Graph_analysis}
\end{figure*}

With the rapid advancement of machine learning and deep learning techniques, an increasing number of studies have explored the use of graph-based learning approaches for IoT malware detection and classification. In these approaches, the analyzed binaries or bytecode are typically transformed into program behavior graphs such as call graphs or control flow graphs (CFGs), while in other cases, device-to-device communication is modeled as a traffic-based graph. In the subsequent stage, these graphs are converted into vector representations: one line of research leverages deep neural networks to perform graph embedding, mapping the entire graph structure into vectors, whereas another line of research relies on feature extraction and selection techniques to compute statistical or structural properties from the graph and construct feature vectors. Finally, during the detection or classification phase, the resulting vectors are fed into trained models to identify or classify IoT malware samples. The workflow of the graph learning representation analysis is shown in Figure \ref{fig:Graph_analysis}.



\subsubsection{Control Flow Graph}



Both \cite{alasmary2019analyzing} and \cite{abusnaina2022dlfhmc} employ Radare2 to transform programs into Control Flow Graphs (CFGs) and extract 23 graph features, such as centrality, shortest path, density, and node/edge counts. These features are embedded into a vector space and then fed into models such as LR, SVM, RF, and CNN for malware detection. Furthermore, in \cite{abusnaina2022dlfhmc}, malicious samples are classified into Gafgyt, Mirai, or Tsunami families, while benign samples are passed to a suspicious behavior detection module. This module mines common subgraph patterns for each family using gSpan, ranks them by size, frequency, coverage, and inverse frequency, and selects the top 10,000 patterns per family (30,000 in total). A 30,000-dimensional one-hot feature vector is then generated using the VF2 subgraph isomorphism algorithm and fed into a machine learning model to determine the presence of suspicious behavior; positive cases are forwarded to dynamic analysis or security experts for further investigation.

Furthermore, After extracting CFGs via Radare2. Soteria \cite{alasmary2020soteria} employs Density-Based (DBL) and Level-Based (LBL) labeling to annotate CFGs, conducts random walks to generate path sequences, and applies the n-gram method to extract subsequences, selecting the top 500 features from each labeling scheme to construct a 1000-dimensional vector. This vector is first processed by a five-layer autoencoder, and if classified as malware, 20 additional vectors (10 from DBL and 10 from LBL) are generated, each classified by CNNs whose outputs are aggregated by majority voting; however, Soteria performs no further analysis once a sample is deemed benign.


\subsubsection{Call Graph}


\cite{yumlembam2023iotbased, deng2024transmalde, deng2024mdhe, lei2019evedroid} focus on graph features derived from API call graphs and leverage these representations for malware detection. Specifically, \cite{yumlembam2023iotbased} employs ApkTool with linear regression–based feature selection to parse APKs and extract key APIs, which are used to construct local API graphs that are subsequently merged into a global API graph. From this graph, five centrality features including degree, betweenness, closeness, eigenvector, and PageRank, are computed. These embeddings are further combined with permissions and intents extracted from the Manifest file, and the fused vector is fed into a classifier for benign/malicious classification.

Similarly, both TransMalDE \cite{deng2024transmalde} and MDHE \cite{deng2024mdhe} utilize Androguard together with Gephi to extract function call graphs (FCGs) from Dalvik bytecode, and focus on identifying sensitive API invocations. During training, the maliciousness degree of each sensitive API is precomputed using the TF-IDF method, where the term frequency reflects the proportion of malware samples invoking the API. In the detection phase, a $k$-hop neighborhood subgraph is constructed for each sensitive API, and its maliciousness degree is defined as the sum of the scores of all sensitive APIs it contains. These subgraph features are normalized and encoded into feature vectors. The two approaches differ in the classification stage: TransMalDE directly feeds the subgraph feature vectors into a Transformer model with a Softmax layer for classification, while MDHE fuses the subgraph features with permission features and inputs them into a Capsule Network (CapsNet) classifier for benign/malicious prediction.

Moreover, EveDroid \cite{lei2019evedroid} extract API call sequence from event-level call graphs for malware detection. it first employs FlowDroid to construct event-level call graphs from Android applications, where each event entry point yields an API call sequence. These calls are decomposed into word sequences (including the package name, return type, method name, and parameter list) and embedded into a semantic vector space using doc2vec. K-means clustering assigns the embeddings to semantic categories, and the frequency distribution of these categories forms the feature representation for each event. Concatenating all event features produces the application’s behavioral vector, which is fed into a neural network for benign/malicious classification.

GCDroid \cite{niu2023gcdroid} extends this idea by constructing an APK–APK graph. The system uses ApkTool to extract API call information and builds an APK–API heterogeneous graph, where APK nodes are linked to sensitive API nodes. It then introduces the Graph Compression with Reachable Relation (GCRR) method to transform this heterogeneous graph into a homogeneous APK–APK graph, where edges indicate shared API invocations weighted by their association strength. Node features, such as the number and proportion of sensitive API calls as well as structural metrics, are preserved and combined with the compressed graph. Finally, a Graph Convolutional Network (GCN) is applied for node classification, labeling each APK as benign or malicious.

Compared with API graphs, function call graphs (FCGs) provide broader coverage and offer a more comprehensive representation of a program’s overall behavior. MaGraMal \cite{feng2024unmasking} employs masked graph representation to detect fine-grained behaviors of IoT malware during the lurking stage. It uses Radare2 to extract function call graphs (FCGs), selects the highest-degree node as the center, and determines an adaptive mask radius to filter peripheral nodes, yielding a masked graph. The graph is embedded with graph2vec\cite{narayanan2017graph2vec} and augmented with three structural features (node count, edge count, graph density), forming a $(d+3)$-dimensional vector. A classifier then predicts four behavior categories: persistence, privilege escalation, deception, and defense evasion.

Furthermore, FedMalDE \cite{pei2023knowledge} partitions the function call graph (FCG) into class-level Sliced Graphs (SGs) based on sub-function call relationships, and selects the top-$k$ SGs with the highest semantic relevance as Sliced Subgraphs (SSGs) for detection. In each SG, a node (class) aggregates multiple feature types: parameter features (e.g., number of parameters), sensitive API features, instruction features (e.g., instruction types), and modifier features (e.g., access and non-access modifiers). Edges encode the shortest calling paths between classes. Then FedMalDE employs a word embedding method to project SSG textual information into an embedding space, which is then input into a Subgraph Aggregated Capsule Network (SACN). A Sigmoid classifier produces the final prediction of whether the sample is malicious or benign.

The printable string information (PSI)-Graph derived from FCGs can accurately capture device-to-device communication patterns. PSI-Graph \cite{nguyen2020novel} enhances FCGs with printable string information (PSIs) extracted from ELF binaries using a customized IDAPython plug-in. PSIs capture critical botnet behaviors such as IP addresses, credentials, and C\&C commands. The resulting PSI-Graph, where nodes are functions containing PSIs and edges denote caller–callee relationships, is embedded with graph2vec into a 1024-dimensional vector and classified by a CNN to detect IoT botnets.

\subsubsection{Dynamic Interaction Graphs}

In contrast to approaches that directly parse binaries or APKs, both Niffler \cite{du2023niffler} and Alharbi et al. \cite{alharbi2021botnet} leverage graph representations to model runtime interactions. Niffler parses event-triggering logic from open-source SmartApp code, uses AstBuilder to build an initial generic correlation graph, infers actuator–sensor dependencies via a Naïve Bayes model, and eliminates orthogonal, irrelevant sensor associations using a Markov chain. Next, by incorporating real-time traffic data, it applies $k$-means clustering (with the Elbow method to select $k$) to refine device grouping, dynamically generating high-precision device-specific correlation graphs. In the detection phase, the dynamic graph stream is represented as a time-series edge set; a biased random walk combined with Node2vec is employed to extract high-sensitivity link paths, where sensitivity is quantified using the Levenshtein distance, retaining only the most likely anomalous interaction paths. Finally, for each device, Niffler builds a local inference model based on the top-$k$ correlated neighbors selected by a GNN and mRMR, predicts the expected device state, and flags it as anomalous if the deviation from the actual state exceeds a predefined threshold.

By comparison, in \cite{alharbi2021botnet}, botnet traffic is modeled as a communication graph using Graph-tool, where nodes represent IP addresses, edges denote communication flows, and weights correspond to packet counts. From this graph, degree-, centrality-, and clustering-based features are extracted, and filter-based selection methods (IG/MI, Gini, correlation, consistency) retain the most relevant ones. The selected features are then classified using multiple supervised models (e.g., NB, DT, RF, AdaBoost, ETC, KNN) to detect and categorize botnet families.

\subsubsection{Summary}

In this section, we first outline the workflow of graph-learning–based analysis methods. We then categorize the literature by graph type including, CFG-based (control-flow graphs), CG-based (call graphs), and DIG-based (dynamic interaction graphs), and discuss the similarities and differences in graph-based feature extraction and malware analysis strategies across these lines of work.

\begin{figure}[htbp]
    \centering
    \includegraphics[width=0.45\textwidth]{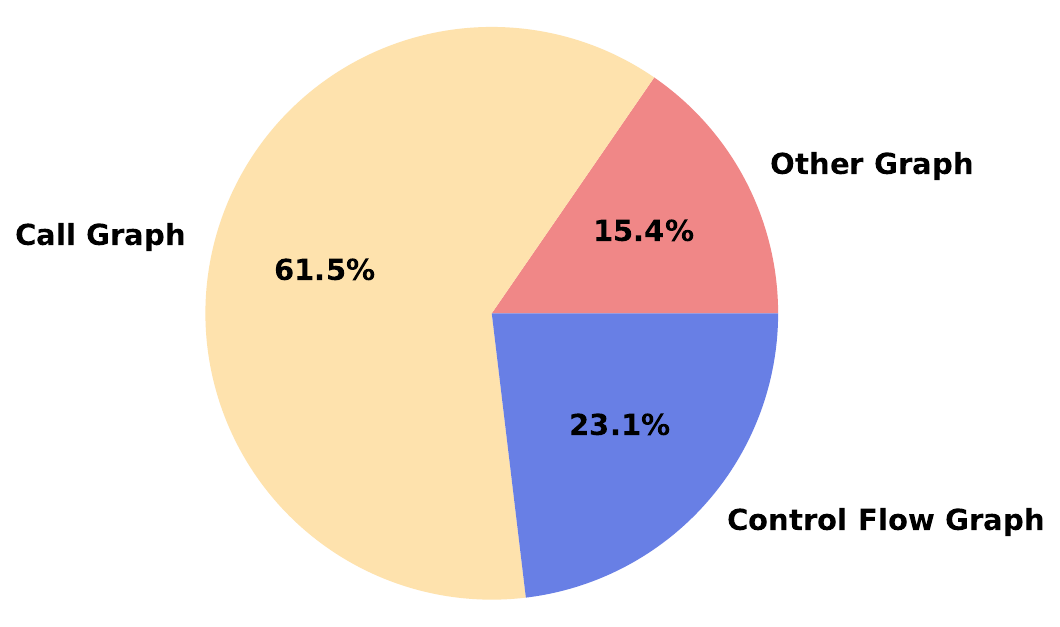}
    \caption{Distribution of feature types for graph-learning-based analysis.}
    \label{fig:graph_types}
\end{figure}

\begin{figure}[htbp]
    \centering
    \includegraphics[width=0.45\textwidth]{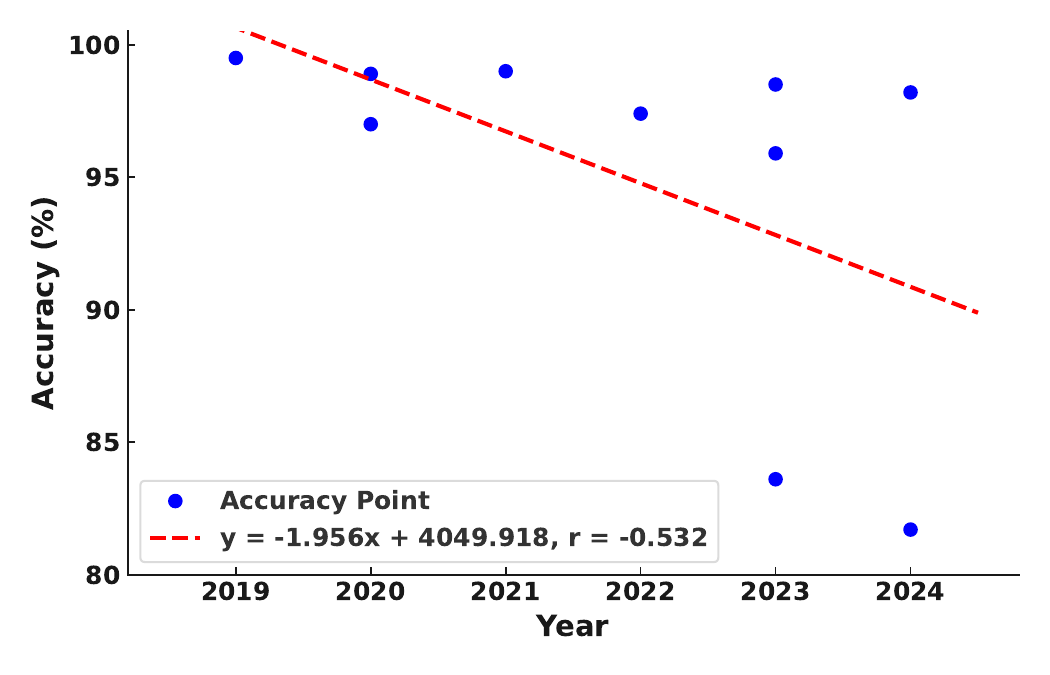}
    \caption{Year vs. Accuracy with OLS regression line for graph-learning-based analysis.}
    \label{fig:graph_learning_regression}
\end{figure}

\begin{itemize}

    \item As shown in Fig.~\ref{fig:graph_types}, call-graph–based methods account for the largest share (61.5\%), while control-flow graphs (CFGs) and dynamic interaction graphs (DIGs) are less common, at 23.1\% and 15.4\%, respectively. Among these graph forms, CFGs offer the finest granularity and highest semantic fidelity, yielding strong potential for precise detection and attribution. However, CFG construction and analysis are time-consuming, which hampers broad adoption in resource-constrained IoT settings—an important reason for their lower share. Call graphs provide coarser granularity and somewhat lower semantic precision than CFGs, yet they can accurately surface sensitive behaviors via sensitive-API invocation relations, subgraph construction, and related techniques, thereby improving analysis accuracy under limited computational budgets; this likely explains their dominant proportion. By contrast, DIGs capture coarse-grained device-to-device behavioral logic and are well suited to IoT deployment, but they suffer from limited temporal persistence and unbounded growth as interactions accumulate, which raises analysis complexity over time and confines their practical use cases.

    \item Using a least-deviation fit to model the relationship between year and method accuracy (see Fig.~\ref{fig:graph_learning_regression}), we find that—after excluding two clear outliers—reported accuracies remain stably within 95\%–100\%. This stability is largely attributable to the maturity of graph construction pipelines and graph-based feature extraction techniques, coupled with the relative homogeneity and closed-world nature of commonly used training and test datasets.
    
    \item Current mainstream methods for feature extraction and embedding can be roughly divided into three categories. The first directly embeds graphs into a vector space using methods such as \texttt{graph2vec}, but the resulting vectors often contain substantial noise, which increases the difficulty of classification or detection for learning-based models. To address this, some studies attempt to extract more representative subgraphs to reduce redundant noise; however, these algorithms are computationally expensive, and balancing subgraph size with analysis efficiency remains an open challenge. The second approach extracts features such as centrality, paths, and node/edge counts from graphs and then maps them into a vector space. This method effectively captures the global properties of graphs while offering lower computational complexity and better interpretability. The third approach combines structural information with semantic contextual information by fusing their vector representations, producing richer features that significantly enhance detection performance.
    
    \item Existing graph learning–based methods are primarily designed for IoT Android or ELF binaries, and their migration to other architectures remains challenging. Meanwhile applying graph-based features for program analysis in IoT environments necessitates careful management of the accuracy–resource trade-off, which remains an ongoing area of exploration and optimization.
\end{itemize}

\section{Opportunities and Challenges}
\label{sec:opptunities and challenges}

 \subsection{Comparison Among Different Analysis Techniques}

Static analysis is a class of lightweight techniques that can accurately locate attacker-delivered malware within IoT systems, addressing attacks at their source. However, current methods infer malware primarily from assembly code and image representations, making them susceptible to obfuscation and cross-architecture issues. In contrast, dynamic analysis can capture the real behavior of IoT devices and is less affected by these problems. Yet it depends on execution environments, and IoT development and security benchmarks are not yet mature, which limits large-scale adoption. Dynamic methods also incur high analysis costs and mostly rely on traffic, while IoT endpoints are resource-constrained and prone to side-channel attacks, further restricting their use. Hybrid analysis can combine the strengths of dynamic and static approaches to improve the accuracy and interpretability of malware inference. To date, hybrid methods include static analysis with dynamic verification, detection by comparing static and dynamic patterns, and joint use of static and dynamic features. These strategies improve the accuracy of single-modality methods but increase overhead and complexity. Graph representation learning offers semantic and structural advantages and strong generalization, but its use involves a trade-off between accuracy and computational cost.

Given the resource constraints of IoT, malware analysis tends to favor lightweight static features. Meanwhile, because IoT attacks are often DDoS and botnets, traffic analysis is widely used in this field. In addition, IoT networks can deploy devices across multiple platforms, leading to strong heterogeneity; this has drawn attention to graph features (e.g., CFGs, API graphs) and cross-platform features. Finally, due to the distributed nature of IoT devices, federated learning and edge computing are widely applied in malware inference.

Throughout the evolution of feature usage and related techniques, we observe that approaches before 2016 mostly relied on simple statistics and single features; starting in 2020, the introduction of deep learning (e.g., BERT, CNNs, LSTMs) improved accuracy; and recent trends emphasize multimodal fusion (semantics + graphs + images) and federated/distributed detection to address IoT resource constraints and privacy concerns.

\subsection{Lessons learned}

After a systematic review of IoT malware analysis, we distill the following key lessons:
\begin{enumerate}

  \item \textbf{No single feature dominates.} Feature fusion is needed to combine complementary strengths and mitigate individual weaknesses.

  \item \textbf{Deployment realities matter.} Resource constraints and platform heterogeneity call for lightweight features with edge/federated designs and cross-ISA evaluation (report latency, memory, and FLOPs).

  \item \textbf{Accuracy improvement has plateaued.} Since 2020, static, dynamic, and graph-learning methods commonly report accuracies in the \(94\%\text{--}98\%\) range. Further gains depend on deep representation learning and multimodal/distributed schemes, along with robust evaluations against obfuscation and cross-architecture shifts.

  \item \textbf{Privacy–utility trade-off.} Federated learning, differential privacy, and distillation/retrieval augmentation are needed to reduce cloud exposure of device data.

  \item \textbf{Standards and benchmarks.} The lack of unified IoT application and security evaluation baselines undermines comparability and real-world deployment.

  \item \textbf{Dataset gaps.} Public datasets are aging and private datasets are unavailable; we recommend time-based splits with leakage control, cross-platform benchmarks, and reproducible protocols.
\end{enumerate}

\subsection{Future Research}

In this survey, we highlight several promising directions for future research. Within traditional static analysis, only Petrache et al. \cite{petrache2025unveiling} employ graph-based features for IoT malware analysis, indicating that this line of research remains underexplored and warrants further investigation. Similarly, hybrid approaches to date have relied almost exclusively on context-sensitive features and have been applied primarily to malware detection rather than classification, leaving classification-oriented hybrid representations as an open avenue. Dynamic analysis, on the other hand, remains heavily dominated by traffic-based features. Future work should explore the integration of traffic with runtime logs, hardware traces, and EM/power spectral features, as well as the construction of graph-structured runtime contexts. Moreover, with the increasing adoption of encrypted protocols (e.g., TLS, DoH), side-channel features and temporal patterns may provide a viable means to recover hidden semantics. Beyond single modalities, robust malware fingerprints may be achieved by fusing static semantics, dynamic behaviors, and physical-layer signals through joint representation learning techniques such as contrastive learning, multimodal transformers, and cross-modal attention.

Given the heterogeneity of IoT ecosystems, malware detection often requires analyzing ELF binaries (ARM/MIPS/x86), Android APKs, and even customized firmware. This motivates the need for cross-architecture and cross-platform features, including ISA-invariant representations (e.g., opcode embedding normalization, obfuscation-invariant signatures) and cross-platform graph features via CFG/API graph embeddings with cross-graph alignment.

Our survey reveals that only 31.9\% of studies fully release their datasets, and among those publicly available, VirusShare, Drebin, IoT-23, Bot-IoT, and VirusTotal together account for approximately 64.5\% of usage. Most of these datasets are outdated, underscoring the necessity of constructing comprehensive benchmarks that span architectures, timeframes, and modalities, along with standardized evaluation protocols to ensure fair and reproducible comparisons.

Finally, the application of large language models (LLMs) to IoT malware analysis remains largely unexplored. LLMs could enhance feature engineering and detection by assisting with reverse engineering, annotating program graphs, and leveraging retrieval-augmented generation (RAG) with malware knowledge graphs to construct semantically enriched features.

\section{CONCLUSION}
\label{sec:conclusion}

Feature selection and extraction techniques form the cornerstone of IoT malware analysis, as effective feature engineering can significantly enhance the accuracy and robustness of malware detection and classification. A key challenge in this field lies in identifying the most effective features and corresponding analysis methods across diverse IoT environments. Addressing this challenge requires continuous investigation into the performance of different feature extraction techniques, as well as the design and optimization of combinations of features with various malware analysis approaches. In recent years, researchers have explored a wide range of such combinations to improve detection and classification outcomes. In this survey, we systematically categorize, compare, and analyze these approaches, and summarize the lessons learned. Furthermore, we release the dataset we collected and propose potential future feature directions to guide subsequent research in this domain.



%
{\footnotesize \bibliographystyle{unsrt}}
\bibliography{Transactions-Bibliography/example}

\end{document}